\documentstyle[epsfig,12pt]{article}
\begin{document}
\title{ Systematic study of the effect of short range correlations
on the form factors and densities of s--p and s--d shell nuclei}

\author{S.E. Massen, H.C. Moustakidis\\
\\
Department of  Theoretical Physics, Aristotle University of Thessaloniki\\
GR-54006 Thessaloniki, Greece}
\maketitle
\begin{abstract}
Analytical expressions of the one-- and two--body terms in the cluster
expansion of the charge form factors and densities of the s--p and s--d shell
nuclei with N$=$Z are derived.
They depend on the harmonic oscillator parameter $b$ and the
parameter $\beta$ which originates from the Jastrow correlation function.
These expressions are used for the systematic study of the effect of
short range correlations on the form factors and densities and of the mass
dependence of the parameters $b$ and $\beta$.
These parameters have been determined by fit to the experimental charge
form factors. The inclusion of the  correlations reproduces the experimental
charge form factors at the high momentum transfers ($q\geq 2 \ $fm$^{-1}$).
It is found that while the parameter $\beta$ is almost constant for the
closed shell nuclei, $^4$He, $^{16}$O and $^{40}$Ca, its values are larger
(less correlated systems) for the open shell nuclei, indicating a shell
effect in the closed shell nuclei.
\\
\\
{PACS numbers: 21.10.Ft, 25.30.Bf, 21.45.+v, 21.60.Cs}
\end{abstract}

\section{INTRODUCTION}

The calculation of the  charge form factors, $F_{ch}(q)$, and density
distributions, $\rho_{ch}(r)$ of nuclei, is a challenging and appealing
problem \cite{Elton67}.
A possibility to face this problem is by means of an independent
particle model. This approach, which is particularly attractive
because of its simplicity, fails to reproduce the high momentum
transfer data from electron scattering in nuclei
\cite{Czyz67,Khana68,Ciofi68,Ciofi70,Ripka70,Gaudin71,Bohigas80,%
DalRi82,Nassena81,Stoitsov93}.
For this reason a modification of the single particle (SP) potentials
has to be suitably made.

In fact a short range repulsion in this potential seems advisable for
light nuclei \cite{Grypeos89}.
For example, with an harmonic oscillator (HO) potential having in addition
an infinite soft core, the $F_{ch}(q)$ of
$^4He$ can be well reproduced, but for the heavier nuclei, such as $^{12}$C
and $^{16}$O, state dependent potentials seem necessary and even then the
fit is not so good for higher q--values \cite{Grypeos89}.
Another way is the introduction of a modified shell model with
fractional occupation numbers for the various states
\cite{Brown79,Malaguti82,Gulkarov,Kosmas92}.

An approach, which is rather similar, is the introduction of the short
range correlations (SRC) in the Slater determinant.
Many attempts have been made in this
direction, concerning mainly light closed shell nuclei in the framework
of the Born--approximation
\cite{Ciofi68,Ciofi70,Ripka70,Gaudin71,Bohigas80,DalRi82,Nassena81,%
Stoitsov93,Massen88,Massen89,Massen90}.

Czyz and Lesniak \cite{Czyz67}  first showed that the
diffraction character of the
$^{4}$He form factor can be qualitatively explained by means of
Jastrow--type \cite{Jastrow55} correlations.
Khana \cite{Khana68}, using Iwamoto--Yamada cluster expansion
and retaining only one-- and two--body cluster terms,
showed that the inclusion of short range nucleon--nucleon
correlations provide an adequate description for the known
data on the elastic scattering of electrons by $^{40}$Ca
and make predictions for the behavior of the cross section at
large momentum transfers.
Ciofi Degli Atti, using the "single pair approximation" \cite{Ciofi68} and
the Iwamoto--Yamada cluster expansion \cite{Ciofi70} in s--p shell nuclei,
showed that the elastic electron scattering at high momentum transfers
seems to give a strong indication of the presence of SRC in nuclei.
Bohigas and Stringari \cite{Bohigas80} and Dal Ri et al \cite{DalRi82}
evaluated the effect of SRC on the one-- and two--body densities by
developing a low order approximation (LOA) in the framework of the Jastrow
formalism.
They showed that one--body quantities like the form factor provide an
adequate test for the presence of SRC in nuclei,
which indicates that the independent--particle wave functions
cannot reproduce simultaneously the form factor and the momentum
distribution of a correlated system.
Stoitsov et al \cite{Stoitsov93}
generalized the model of Jastrow  correlations,
suggested by Bohigas and Stringari \cite{Bohigas80} within the LOA of Ref.
\cite{Gaudin71}, to heavier nuclei like $^{16}$O, $^{36}$Ar
and $^{40}$Ca reproducing very well the experimental data.

In the above approaches, different types of expansions were used.
The expansions were connected with the number of
the simultaneously correlated nucleons, and the order of the Jastrow
correlation function, $f(r_{ij})$, which were retained in the cluster
expansion. Usually they are truncated up to the two--body terms which give
significant contribution to various expectation values.

In a series of papers \cite{Massen88,Massen89,Massen90}
an expression of the elastic charge form factor, $F_{ch}(q)$,
truncated at the two--body term, was derived using the factor
cluster expansion of Clark et al
\cite{Clark70,Ristig-Clark,Clark79}.
This expression, which is a sum of one--body and two--body terms,
and depends on the HO parameter and the correlation parameter
through a Jastrow type correlation function, is used for the calculation
of $F_{ch}(q)$ of $^4$He, $^{16}$O and $^{40}$Ca and in an approximate way
for the open s--p and s--d shell nuclei.

The motivation of the present work is the systematic study of the effect
of SRC on the s--p and s--d shell nuclei by completely avoiding the
approximation made in earlier work \cite{Massen88,Massen89,Massen90}
for open shell nuclei, that is the expansion of the two--body terms in
powers of the correlation parameter, in which only the leading terms had
been retained.
General expressions for the $F_{ch}(q)$ and $\rho_{ch}(r)$ were
found using the factor cluster expansion of Clark et al and Jastrow
correlation functions which introduce SRC.
These expressions are functionals of the SP wave functions
and not of the wave functions of the relative motion of two nucleons
as was the case in  many previous works \cite{Ciofi68,Nassena81,Massen88}.
Because of that, it is easy to extrapolate them to the case of open shell
nuclei and use them either for analytical calculations with HO wave
functions or for numerical calculations when more realistic
SP wave functions are used.
An advantage of the present method is that the mass dependence of the
HO parameter $b$ (with the presence of correlations) and the correlation
parameter $\beta$ can be studied. These parameters have been determined, for
the various s--p and s--d shell nuclei by fit of the theoretical $F_{ch}(q)$
to the experimental ones.
It is found that while the parameter $\beta$ is almost constant
for the closed shell nuclei, $^4$He, $^{16}$O and $^{40}$Ca, it takes
larger values (less correlated systems) in the open shell nuclei,
indicating a shell effect for the closed shells.
The present method has also been used with
wave functions of  Skyrme--type interactions. In this case
there is only one free parameter, the correlation parameter $\beta$
and the two body--term is a small correction to the one body--term of the
density.

The paper is organized as follows. In Section II, the general
expressions of the correlated form factors and  density distributions are
derived using a Jastrow correlation function. In section III
the analytical expressions of the above quantities for the s--p and s--d
shell nuclei, in the case of the HO orbitals, are given.
Numerical results are reported and discussed in section IV, while
an outline of the present work is given in section V.

\section{CORRELATED DENSITY DISTRIBUTIONS AND FORM FACTORS}

If we denote the model operator, which introduces SRC, by
$\cal{F}$,
an eigenstate $\Phi$ of the model system corresponds to an eigenstate
\begin{equation}
\Psi={\cal F}\Phi
\label{eq1}
\end{equation}
of the true system.

Several restrictions can be made on the model operator $\cal{F}$,
as for example, that it depends on (the spins, isospins and) relative
co--ordinates and momenta of the particles in the system, it is a scalar
with respect to rotations e.t.c. \cite{Brink67}. Further, it is required that
$\cal{F}$ is translationally invariant and symmetrical in its argument
$1 \cdots i \cdots A$ and possesses the cluster property. That is if any
subset, $i_1 \cdots i_p$, of the particles is removed far from the rest,
$i_{p+1} \cdots i_A$, $\cal{F}$ decomposes into a product of two factors,
${\cal F}(1\cdots A) =
{\cal F}(i_1 \cdots i_p) \ {\cal F}(i_{p+1} \cdots i_A)$ \cite{Clark79}.
In the present work $\cal{F}$ is taken to be of the Jastrow--type
\cite{Jastrow55},
\begin{equation}
{\cal F}=\prod_{i<j}^{A}f(r_{ij})\ ,
\end{equation}
where $f(r_{ij})$ is the state-independent correlation function
of the form:
\begin{equation}
f(r_{ij})=1-\exp[-\beta({\bf r}_i-{\bf r}_j)^2] \ .
\label{fr-ij}
\end{equation}

The correlation function $f(r_{ij})$ goes to one for large values of
$r_{ij} = \mid {\bf r}_i - {\bf r}_j \mid$ and it goes to zero for
$r_{ij} \rightarrow 0$. It is obvious that the effect of SRC, introduced
by the function $f(r_{ij})$, becomes large when the SRC parameter
$\beta$ becomes small and vice versa.

The charge form factor of a nucleus, in Born--approximation, can be written
\begin{equation}
F_{ch}(q)  = f_{p}(q)\ f_{DF}(q) \ f_{CM}(q) \  F_{p}(q) \ ,
\label{Fch-1}
\end{equation}
where $f_{p}(q)$  and $f_{DF}(q)$ are the correction for the finite proton
size and the Darwin--Foldy relativistic correction, respectively
\cite{Chandra76}, $f_{CM}(q)$ is the Tassie--Barker \cite{Tassie58}
center--of--mass correction and $F_p(q)$ the point form factor of the
nucleus which is the expectation value of the one--body operator,
\begin{equation}
{\bf O}_q \ =\ \sum_{i=1}^{A} {\bf o}_{q}(i)\
= \ \sum_{i=1}^{A} \exp [i{\bf q} \ {\bf r}_i] \ .
\label{O-q}
\end{equation}
That is,
\begin{equation}
F_{p}(q)\ = \frac{\langle\Psi|{\bf O}_q|\Psi\rangle}{\langle\Psi|\Psi\rangle}
\ =\ N \langle\Psi|{\bf O}_q|\Psi\rangle \ =\ N \langle {\bf O}_q \rangle
\ ,
\label{F-p1}
\end{equation}
where
$N=\langle\Psi(r_1,r_2,\cdots ,r_A)|\Psi(r_1,r_2,\cdots ,r_A) \rangle^{-1}$
is the normalization factor which
is determined so that $F_{ch}(0) = F_{p}(0) = 1 $ or
$4 \pi \int_0^\infty \rho (r) r^2 {\rm d} r = 1$.

The point density distribution has the form
\begin{equation}
\rho_p(r)=\frac{\langle \Psi|{\bf O}_r|\Psi\rangle}{\langle\Psi|\Psi\rangle}\
 =\ N \langle \Psi|{\bf O}_{r}|\Psi\rangle \
=\ N \langle {\bf O}_{r} \rangle \ ,
\label{den-point}
\end{equation}
where
\begin{equation}
{\bf O}_{r} \ =\ \sum_{i=1}^{A} {\bf o}_{r}(i)\
=\ \sum_{i=1}^{A}\delta({\bf r}-{\bf r}_i) \ .
\label{O-r}
\end{equation}

\subsection{EXPRESSIONS OF THE CORRELATED DENSITY DISTRIBUTIONS
AND  FORM FACTORS \label{sub2.1}}
In order to evaluate the point density distribution,
$\rho_{p}(r)$, we consider, first, the generalized normalization integral,
\begin{equation}
I(\alpha)=\langle\Psi | \exp [\alpha I(0){\bf O}_r] |\Psi\rangle \ ,
\label{I-1}
\end{equation}
corresponding to the operator ${\bf O}_r$, from which we have,
\begin{equation}
\langle{\bf O}_r\rangle =
\left[ \frac{\partial \ln I (\alpha)}
{\partial \alpha} \right]_{\alpha=0} .
\label{I-2}
\end{equation}

For the cluster analysis of equation (\ref{I-2}),
following the factor cluster expansion of Ristig and Clark
\cite{Clark70,Ristig-Clark,Clark79}, we consider the sum--product
integrals, $I_{i}(\alpha),\ I_{ij}(\alpha),\cdots$, for the subsystems of the
A--nucleon system and a factor cluster decomposition of these integrals.
The expectation value of the density distribution operator is
written in the form,
\begin{equation}
\langle {\bf O}_r \rangle=\langle {\bf O}_r \rangle_1 +
\langle{\bf O}_r\rangle_2 + \cdots + \langle {\bf O}_r \rangle_{A} \ ,
\end{equation}
where
\begin{eqnarray}
\langle {\bf O}_r \rangle _1 &=&
 \sum_{i=1}^{A} \left[ \frac{\partial \ln I_{i}(\alpha)}{\partial \alpha}
\right]_{\alpha=0} \ = \
 \sum_{i=1}^{A} \langle i \mid {\cal F}_1^\dag {\bf o}_{r}(1){\cal F}_1
\mid i \rangle \ ,
\label{Or1-1}
\end{eqnarray}
\begin{eqnarray}
\langle {\bf O}_r \rangle_2 &=&
\sum_{i<j}^{A} \frac{\partial}{\partial \alpha} \left[
 \ln I_{ij}(\alpha) - \ln I_i(\alpha) -  \ln I_j(\alpha) \right]_{\alpha=0}
\nonumber\\
&=& \sum_{i<j}^{A}\langle ij \mid {\cal F}_{12}^\dag ({\bf o}_{r}(1)
+{\bf o}_{r}(2))
{\cal F}_{12}\mid ij \rangle_a -
 \sum_{i<j}^{A} [ \langle i \mid {\bf o}_{r}(1)\mid i \rangle +
\langle j \mid {\bf o}_{r}(2) \mid j \rangle ] \ ,
\label{Or2-1}
\end{eqnarray}
and so on. ${\cal F}_1$ is chosen to be the identity operator.

The cluster expansion establishes the separation of one--body, two--body,
$\cdots$, A--body correlation effects on the density. Three-- and
many--body terms will be neglected in the present analysis. Thus, in
the two--body approximation, writing also the two sums of equation
(\ref{Or2-1}) in a different form, $\rho_{p}(r) $ is written as follows,
\begin{equation}
\rho_{p}(r)=N  \langle {\bf O}_r \rangle
\approx N [ \langle {\bf O}_r \rangle_1
+\langle {\bf O}_r \rangle_{22} - \langle {\bf O}_r \rangle_{21} ] \ ,
\label{Dp-2}
\end{equation}
where
\begin{eqnarray}
\langle {\bf O}_r \rangle _1 &=&
 \sum_{i=1}^{A} \langle i \mid {\bf o}_{r}(1) \mid i \rangle \ ,
\label{Or1-2}\\
\langle {\bf O}_r \rangle_{22} &=&
2 \sum_{i<j}^{A}\langle ij \mid {\cal F}_{12}^{\dag} {\bf o}_{r}(1)
{\cal F}_{12}\mid ij \rangle_a \ ,
\label{Or22-1}\\
\langle {\bf O}_r \rangle_{21} &=&
2 \sum_{i<j}^{A}  \langle ij \mid  {\bf o}_{r}(1) \mid ij \rangle_a \ .
\label{Or21-1}
\end{eqnarray}

If the  two--body operator is taken to be the correlation function given by
(\ref{fr-ij}), then
\begin{equation}
{\cal F}_{12}^{\dag} {\cal F}_{12} =
1 - 2 g(r_1,r_2,\beta) + g(r_1,r_2,2\beta)\ ,
\label{F12F12}
\end{equation}
where
\begin{equation}
 g(r_1,r_2,z) = \exp [-z r_1^2] \exp [-z r_2^2]
\exp [2 z r_1 r_2 \cos \omega_{12}]\ , \quad z=\beta, 2\beta \ ,
\label{gr12}
\end{equation}
and the term $\langle {\bf O}_r \rangle_{22}$ is written,
\begin{equation}
\langle {\bf O}_r \rangle_{22} =\langle {\bf O}_r \rangle_{21}
- 2 O_{22}(r,\beta) + O_{22}(r,2 \beta) \ ,
\label{O22-2}
\end{equation}
where
\begin{equation}
O_{22}(r,z) = 2 \sum_{i<j}^{A}  \langle ij \mid  {\bf o}_{r}(1) g(r_1,r_2,z)
\mid ij \rangle_a \ .
\label{O22-g-1}
\end{equation}
So, $\rho_{p}(r)$ takes the form,
\begin{equation}
\rho_{p}(r) \approx N [ \langle {\bf O}_r \rangle_1 - 2 O_{22}(r,\beta) +
O_{22}(r,2 \beta)] \ .
\label{Dp-3}
\end{equation}

The terms $\langle {\bf O}_r \rangle_1$ and $O_{22}(r,z)$ and the density
$\rho_{p}(r)$ can be expressed also in the convenient form:
\begin{equation}
\langle {\bf O}_r \rangle_1=\rho_{SD}({\bf r})\ ,
\end{equation}
\begin{equation}
O_{22}(r,z)=\int g(r,r_2,z)[\rho_{SD}({\bf r})\rho_{SD}({\bf r}_2)-
\rho_{SD}^2({\bf r},{\bf r}_2)] {\rm d}{\bf r}_2 \ ,
\end{equation}
\begin{equation}
\rho_{p}(r) \approx N \left[\rho_{SD}({\bf r})+\int
\left[g(r,r_2,2\beta)-2g(r,r_2,\beta) \right]
\left[\rho_{SD}({\bf r})\rho_{SD}({\bf r}_2)-
\rho_{SD}^2({\bf r},{\bf r}_2)\right] \right] {\rm d}{\bf r}_2 \ ,
\label{Dp-4}
\end{equation}
where $\rho_{SD}({\bf r}_1,{\bf r}_2)$ is the uncorrelated density matrix
associated with the Slater determinant,
\begin{equation}
\rho_{SD}({\bf r}_1,{\bf r}_2)=\sum_{i=1}^{A}\phi_i^*({\bf r}_1)
\phi_i({\bf r}_2) \ .
\end{equation}
The diagonal elements of this gives the one body density
distribution,
\begin{equation}
\rho_{SD}({\bf r}_1)= \rho_{SD}({\bf r}_1,{\bf r}_1) \ .
\end{equation}

It should be noted that, a similar expression for $\rho_p(r)$, given by
equation (\ref{Dp-4}), was derived by Gaudin et al. \cite{Gaudin71}
in the framework  of LOA. In LOA the Jastrow wave function, $\Psi$,
of the nucleus was expanded in terms of the functions:
$\tilde{g}=f^2(r_{ij})-1$ and $h=f(r_{ij})-1$ and was truncated up to
the second order of $h$ and the first order of $\tilde{g}$. This expansion
contains one-- and two--body terms and a part of the three--body term
which was chosen so that the normalization of the wave function was
preserved. Expression (\ref{Dp-4}) of the present work has only one-- and
two--body terms and the normalization of the wave function is preserved
by the normalization factor $N$.

In the above expression of $\rho_p(r)$, the one--body contribution to the
density is well known and given by the equation,
\begin{equation}
\langle {\bf O}_r \rangle_1 =4 \sum_{nl} \eta_{nl} (2l+1)
\frac{1}{4\pi} \phi^{*}_{nl}(r) \phi_{nl}(r) \ ,
\label{O1-3}
\end{equation}
where $\eta_{nl}$ is the occupation probability of the state $nl$ (0 or 1
in the case of closed shell nuclei) and
$\phi_{nl}(r)$ is the radial part of the SP wave function.

An expression for the two--body term is usually found by making a
transformation to the relative and the center--of--mass coordinates of the
two interacting nucleons \cite{Ciofi68,Nassena81,Massen88}.
This is because the Jastrow function $f(r_{ij})$ depends on the
relative coordinates of the two nucleons.
Here, the expression for the two--body term, that is of the term $O_{22}(r,z)$,
will be found by expanding the factor $\exp [2 z r_1 r_2 \cos \omega_{12}]$
in spherical harmonics \cite{Roy,deSalit}. That is
\begin{equation}
 \exp [2 z r_1 r_2 \cos \omega_{12}] = 2 \pi
\sum_{k m_k} U_k( 2 z r_1 r_2)
 Y_{km_k}^{*}(\Omega_1) Y_{km_k}(\Omega_2) \ ,
\label{e-cos}
\end{equation}
where
\begin{eqnarray}
U_k( 2 z r_1 r_2)& =& \int_{-1}^1 \exp (2 z r_1 r_2 \cos \omega_{12})
P_k(\cos \omega_{12}) {\rm d} (\cos \omega_{12}) \nonumber \\
\nonumber \\
& =& 2 i_k (2z r_1 r_2)\ ,
\label{U-k}
\end{eqnarray}
$i_k(x)$ is the modified spherical Bessel function.

Using the algebra of spherical harmonics, the term:
$O_{22}(r,z)$ takes the form:
\begin{eqnarray}
O_{22}(r,z)& =& 4 \sum_{n_i l_i,n_j l_j} \eta_{n_i l_i} \eta_{n_j l_j}
(2 l_i +1) (2 l_j +1 ) \times \nonumber \\
& &\left[
4 A_{n_il_in_jl_j}^{n_il_i n_jl_j k}(r,z)
- \sum_{k=0}^{l_i +l_j}
\langle l_i 0 l_j 0 \mid k0 \rangle^2
 A_{n_il_in_jl_j}^{n_jl_j n_il_i k}(r, z) \right] ,
\quad z=\beta, 2\beta \ ,
\label{O22-g-3}
\end{eqnarray}
where
\begin{eqnarray}
A_{n_1l_1n_2l_2}^{n_3l_3n_4l_4k}(r,z)& =&
\frac{1}{4\pi} \phi^{*}_{n_1l_1}(r)
\phi_{n_3l_3}(r)\exp[-z r^2] \times \nonumber \\
& &\int_{0}^{\infty}\phi^{*}_{n_2l_2}(r_2)\phi_{n_4l_4}(r_2)
\exp[-z r_{2}^2] i_k (2 z r r_2) r_{2}^{2} {\rm d}r_2 \ ,
\label{A-O22-1}
\end{eqnarray}

Thus the expression of the term $O_{22}(r,z)$  depends on the SP
wave functions and so it is suitable to be used either for analytical
calculations with the HO potential or for numerical calculations with
more realistic SP potentials.
Expressions (\ref{O1-3}) and (\ref{O22-g-3}) were derived for the closed
shell nuclei with N$=$Z, where $\eta_{nl}$ is 0 or 1. For the open shell
nuclei (with N$=$Z) we use the same expressions, where now:
$0 \le \eta_{nl} \le 1$. In this way the mass dependence of the
correlation parameter $\beta$ and the HO parameter $b$ can be studied.

Finally, using the known values of the Clebsch--Gordan coefficients and
provided that $\phi_{nl}^{*}(r)=\phi_{nl}(r)$, i.e.:
$A_{n_1l_1n_2l_2}^{n_3l_3n_4l_4k}(r,z)=
A_{n_3l_3n_4l_4k}^{n_1l_1n_2l_2}(r,z)$,
equation (\ref{O22-g-3}), for the case of s--p and s--d shell nuclei,
takes the form:
\begin{eqnarray}
O_{22}(r,z)&=&4\left[ 3A_{0000}^{00000}(r,z)\eta_{1s}^{2}
+\left(33 A_{0101}^{01010}(r,z) - 6 A_{0101}^{01012}(r,z) \right) \eta_{1p}^2
+3A_{1010}^{10100}(r,z)\eta_{2s}^2 \right. +
\nonumber \\
& &\left( 95 A_{0202}^{02020}(r,z) - \frac{50}{7}A_{0202}^{02022}(r,z)
-\frac{90}{7}A_{0202}^{02024}(r,z) \right) \eta_{1d}^2 +
\nonumber \\
& &\left( 12A_{0001}^{00010}(r,z)+12A_{0100}^{01000}(r,z)
-6A_{0001}^{01001}(r,z)  \right) \eta_{1s} \eta_{1p} +
\nonumber\\
& &\nonumber\\
& &\left(20A_{0002}^{00020}(r,z)+20A_{0200}^{02000}(r,z)
-10A_{0002}^{02002}(r,z)  \right) \eta_{1s} \eta_{1d} +
\nonumber\\
& &\nonumber\\
& & \left( 60A_{0102}^{01020}(r,z) + 60A_{0201}^{02010}(r,z)
-12A_{0102}^{02011}(r,z) -18A_{0102}^{02013}(r,z)  \right)
\eta_{1p} \eta_{1d} +   \nonumber\\
& &\nonumber\\
& &\left( 4A_{0010}^{00100}(r,z)+4A_{1000}^{10000}(r,z)
-2 A_{0010}^{10000}(r,z)  \right) \eta_{1s} \eta_{2s} +
\nonumber\\
& &\nonumber\\
& &\left( 12A_{0110}^{01100}(r,z) + 12A_{1001}^{10010}(r,z)
-6A_{0110}^{10011}(r,z) \right)\eta_{1p} \eta_{2s} +
\nonumber\\
& &\nonumber\\
& &\left. \left( 20A_{0210}^{02100}(r,z)+20A_{1002}^{10020}(r,z)
-10A_{0210}^{10022}(r,z)  \right) \eta_{_{1d}} \eta_{2s}
 \right] \ .
\label{O22-A}
\end{eqnarray}

The point form factor $F_{p}(q)$ can be derived in two equivalent
ways. The first one is to follow the same cluster expansion as in the case of
the density distribution and the second one is to take the Fourier transform
of the density distribution  $\rho_{p}(r)$,
\begin{equation}
F_{p}(q)=\int \exp [i{\bf q} \ {\bf r}] \rho_{p}(r) {\rm d}{\bf r} \ .
\label{F-T}
\end{equation}
In both cases, the form factor takes the following form:
\begin{equation}
F_{p}(q) \approx N [ \langle {\bf O}_q \rangle_1 -
2 \tilde{O}_{22}(q,\beta) + \tilde{O}_{22}(q, 2 \beta)] \ .
\label{Fp-3}
\end{equation}
In the above expression, the one--body term is given by the equation
\begin{equation}
\langle {\bf O}_q \rangle_1 =4 \sum_{nl} \eta_{nl} (2l+1)
\int_{0}^{\infty}\phi^{*}_{nl}(r) \phi_{nl}(r) j_{0}(qr)r^{2} dr \ ,
\end{equation}
while the two--body term $\tilde{O}_{22}(q,z)$ is given by the right hand
side of equations (\ref{O22-g-3}) and (\ref{O22-A}) by replacing the
matrix elements $A_{n_1l_1n_2l_2}^{n_3l_3n_4l_4k}(r,z)$ by
$\tilde{A}_{n_1l_1n_2l_2}^{n_3l_3n_4l_4k}(q,z)$ given by the equation,
\begin{eqnarray}
\tilde{A}_{n_1l_1n_2l_2}^{n_3l_3n_4l_4k}(q, z)&=&
\int_{0}^{\infty}\phi^{*}_{n_1l_1}(r_1)
\phi_{n_3l_3}(r_1) \exp[- z r_1^2] j_0(qr_1) r_1^2 {\rm d} r_1 \times
\nonumber\\
& &\int_{0}^{\infty}\phi^{*}_{n_2l_2}(r_2)\phi_{n_4l_4}(r_2)
\exp[- z r_2^2] i_k(2 z r_1r_2) r_2^2 {\rm d} r_2 \ ,
\label{A-O22-2}
\end{eqnarray}
where $j_0(x)$ and $i_k(x)$ are the spherical Bessel and the modified
spherical Bessel functions respectively.


\subsection{l--s COUPLING SCHEME\label{sub2.2}}

In the cases of mean field potentials with l--s coupling,
the wave function of the $nlj$ state has the form:
\begin{equation}
\psi_{nlj}(r_1)=\phi_{nlj}(r_1)
\sum_{m_{l},m_{s}} \langle lm_lsm_s|jm_j\rangle  Y_{l}^{m_l}(\Omega)
\chi_{s}^{m_s}(1)  \chi_{t}^{m_t}(1) \ ,
\end{equation}
where $\phi_{nlj}(r_1)$ is the radial part of the wave function of the
state $nlj$ and $\chi_{s}^{m_s}(1)$, $\chi_{t}^{m_t}(1)$ are the spin and the
isospin wave functions respectively.

Following the same procedure as in sub--section \ref{sub2.1},
the same expression (\ref{Dp-3}) for $\rho _p(r)$ has been found.
In this case the terms $\langle {\bf O}_r \rangle_1$ and $O_{22}(r,z)$
have the form:
\begin{equation}
\langle {\bf O}_r \rangle_1 =2 \sum_{nl} \eta_{nlj} (2j+1)
\frac{1}{4\pi} \phi_{nlj}^{*}(r) \phi_{nlj}(r)
\end{equation}
and
\begin{eqnarray}
O_{22}(r,z)& =& 2 \sum_{n_i l_i j_i,n_j l_j j_j}
\eta_{n_i l_i j_i} \eta_{n_j l_j j_j}
(2 j_i +1) (2 j_j +1 ) \times \nonumber \\
& &\left[2 A_{n_il_ij_in_jl_jj_j}^{n_il_ij_i n_jl_jj_j k}(r, z)
-\frac{1}{2} \sum_{k=0}^{l_i +l_j}
\langle l_i 0 l_j 0 \mid k0 \rangle^2
 A_{n_il_ij_in_jl_jj_j}^{n_jl_j j_jn_il_ij_i k}(r, z) \right]\ ,
 \quad z=\beta, 2 \beta \ ,
\label{O22-g-4}
\end{eqnarray}
where the matrix elements $A_{n_1l_1j_1n_2l_2j_2}^{n_3l_3j_3n_4l_4j_4k}(r,z)$
are given again by equation (\ref{A-O22-1}) replacing the wave
functions $\phi_{nl}(r)$ by the wave functions $\phi_{nlj}(r)$.
Equation (\ref{O22-g-4}) can also be written in a similar form with that
of equation (\ref{O22-A}) with the difference that there are now more terms.
This expression goes to (\ref{O22-A}) if there is no l--s coupling.

In the evaluation of the density matrix $\rho_{SD}({\bf r}_1,{\bf r}_2)$,
which is necessary for the derivation of the expression (\ref{O22-g-4}) of
the term $O_{22}(r,z)$, we made the
following approximation: In the sum over the spin coordinates, only terms
of pairs of particles having the same third spin component are taken into
account. The contribution of the  terms which contain pairs with opposite
third spin component is small and is neglected \cite{Arias96,Arias97}.
In this scheme the form factor is calculated numerically by Fourier
transform of $\rho_p(r)$ employing equation (\ref{F-T}).

\section{ANALYTICAL EXPRESSIONS\label{sec-3} }

In the case of the HO wave functions, with radial part,
\begin{equation}
\phi_{nl}(r)=N_{nl}b^{-3/2}  \left(\frac{r}{b}\right)^l
L_n^{l+\frac{1}{2}} \left( \frac{r^2}{b^2} \right) \
\exp\left[- \frac{r^2}{2 b^2} \right] \ , \qquad
N_{nl}=\left(\frac{2 n!}{\Gamma(n+l+\frac{3}{2})} \right)^{1/2}
\ ,
\end{equation}
where the Associated Laguerre polynomial is defined as in Ref.
\cite{Grads},
analytical expressions of the one--body term and of the matrix elements
$A_{n_1l_1n_2l_2}^{n_3l_3n_4l_4 k}(r,z)$
and $\tilde{A}_{n_1l_1n_2l_2}^{n_3l_3n_4l_4 k}(q,z)$,
defined by equation (\ref{A-O22-1}) and (\ref{A-O22-2}), can be found.
From these expressions, the analytical expressions of the terms
$O_{22}(r,z)$ and $\tilde{O}_{22}(q,z)$, defined by equation
(\ref{O22-A}), can also be found.

The expression of the one--body term of the density and form factor
has the form:
\begin{equation}
\langle {\bf O}_{x} \rangle_1=C \ {\rm e}^{-\xi ^2} \
\sum_{k=0}^{2}C_{2k}\xi^{2k} \ ,\quad x=r,q \ ,
\end{equation}
where for the $\langle{\bf O}_{r}\rangle_1$ the variable $\xi$
and the coefficients $C$ and $C_{2k}$ are:
\begin{eqnarray}
\xi&=&\frac{r}{b} \ , \qquad C=\frac{2}{\pi^{3/2} b^3} \ , \nonumber\\
C_0&=&2\eta_{1s}+3\eta_{2s} \ , \nonumber\\
C_2&=& 4(\eta_{1p}-\eta_{2s}) \ , \\
C_4&=& \frac{4}{3} (2 \eta_{1d}+  \eta_{2s}) \ , \nonumber
\end{eqnarray}
while for the $\langle{\bf O}_{q}\rangle_1$ the corresponding quantities
are:
\begin{eqnarray}
\xi&=&\frac{1}{2}bq\ , \qquad  C=2 \ , \nonumber\\
C_0&=&2( \eta_{1s} + \eta_{2s} + 3 \eta_{1p} + 5 \eta_{1d}) \ , \nonumber\\
C_2&=&-\frac{4}{3} (3\eta_{1p}+10\eta_{1d} + 2\eta_{2s}) \ ,\\
C_4&=&\frac{4}{3} (2\eta_{1d} + \eta_{2s})\ . \nonumber
\end{eqnarray}

The analytical expression of the matrix element
$A_{n_1l_1n_2l_2}^{n_3l_3n_4l_4k}(r,z)$, which are given by
equation (\ref{A-O22-1}), has the form:
\begin{eqnarray}
A_{n_1l_1n_2l_2}^{n_3l_3n_4l_4k}(r,z)&=&
\left( \prod_{i=1}^{4}  N_{n_il_i} \right)
\frac{y^{k}}{16  \sqrt{\pi} b^3 }
\  \xi^{l_1+l_3+k} \ L_{n_1}^{l_1+\frac{1}{2}}(\xi^2) \
L_{n_3}^{l_3+\frac{1}{2}}(\xi^2)
\exp \left[- \frac{1+2y}{1+y}\xi^2 \right] \times \nonumber\\
&& \nonumber\\
&&
\sum_{w=0}^{n_2}\sum_{s=0}^{n_4}
\frac{(-1)^{w+s}}{w!s!}
\left(  \begin{array}{c}
n_{2}+l_{2}+\frac{1}{2} \\
n_{2}-w
\end{array}  \right)
\left(  \begin{array}{c}
n_{4}+l_{4}+\frac{1}{2} \\
n_{4}-s
\end{array}  \right)
  \times \nonumber\\
&&  \nonumber\\
&&
\frac{(\frac{1}{2}(l_2+l_4-k)+w+s)!}
{(1+y)^{\frac{1}{2}(l_2+l_4+k+3) w+s}}  \
L_{\frac{1}{2}(l_2+l_4-k)+w+s}^{k+\frac{1}{2}}
 \left(\frac{-y^2}{1+y} \xi^2 \right)
\label{A-anal-r}
\end{eqnarray}
where $\xi = r/b$ and $y=z b^2$ ($z=\beta,\ 2\beta$).
It is mentioned that the Clebsch--Gordan coefficients
$\langle l_i 0 l_j 0|k0\rangle $, which appear in equation (\ref{O22-g-3})
(where $k$ runs from $0$ to $l_i + l_j$), are different from zero only
when $l_i +l_j +k$ is an even integer and therefore  $l_2 +l_4 - k$ is
an even integer. Because of that the lower index of the Laguerre
polynomial of equation (\ref{A-anal-r}) is an integer.

The above expression of
$A_{n_1l_1n_2l_2}^{n_3l_3n_4l_4k}(r,z)$ is of the form:
$f(\xi^2) \exp [- \frac{1+2y}{1+y} \xi^2]$, where $f(\xi^2)$ is a polynomial
of $\xi^2$.
The substitution of the expression of
$A_{n_1l_1n_2l_2}^{n_3l_3n_4l_4k}(r,z)$
to the expression of $O_{22}(r,z)$, which is given by equation
(\ref{O22-A}), leads to the analytical expression of the two--body
term of the density. This expression, for the case
of s--p and s--d shell nuclei, is again of the form
$f(\xi^2) \exp [- \frac{1+2y}{1+y}\xi^2]$ where now $f(\xi^2)$ is
a polynomial of the form:
\begin{equation}
f(\xi^2)=\sum_{k=0}^{4} C_{2k}(\eta_{n_k l_k},y)  \xi^{2k}\ .
\label{f-xi2}
\end{equation}

The corresponding analytical expression of the matrix element
$\tilde{A}_{n_1l_1n_2l_2}^{n_3l_3n_4l_4k}(q,z)$, which are given by
equation (\ref{A-O22-2}), has the form:
{\small
\begin{eqnarray}
\tilde{A}_{n_1l_1n_2l_2}^{n_3l_3n_4l_4k}(q,z)&=&
\frac{\pi}{16} \left( \prod_{i=1}^{4}  N_{n_il_i} \right)
\exp\left[-\frac{1+y}{1+2y} \xi^2 \right]
\sum_{p=0}^{n_1}\sum_{{\rm v}=0}^{n_3}\sum_{w=0}^{n_2}\sum_{s=0}^{n_4}
\sum_{t=0}^{\frac{1}{2}(l_2+l_4-k)+w+s}
\frac{(-1)^{p+{\rm v}+w+s+t}}{p!{\rm v}!w!s! t!}
 \times
\nonumber\\
&&\nonumber\\
&&
\left(
\begin{array}{c}
n_{1}+l_{1}+\frac{1}{2} \\
n_{1}-p
\end{array}   \right)
\left(   \begin{array}{c}
n_{3}+l_{3}+\frac{1}{2} \\
n_{3}-{\rm v}
\end{array}   \right)
\left(    \begin{array}{c}
n_{2}+l_{2}+\frac{1}{2} \\
n_{2}-w
\end{array}   \right)
\left(   \begin{array}{c}
n_{4}+l_{4}+\frac{1}{2} \\
n_{4}-s
\end{array}  \right)
 \times  \nonumber\\
&& \nonumber\\
&&
\frac{
\Gamma\left(\frac{1}{2}(l_2+l_4+k+3)+w+s\right)
\left(\frac{1}{2}(l_1+l_3+k)+p+{\rm v}+t \right)!
\left(\frac{1}{2}(k-l_2-l_4)-w-s \right)_{t}
}
{\Gamma\left(t+k+\frac{3}{2} \right)
}  \times \nonumber\\
&& \nonumber\\
&&
y^{2t +k} \
\frac{
(1+y)^{\frac{1}{2}(l_1+l_3-l_2-l_4)+p+{\rm v}-w-s}}
{(1+2y)^{\frac{1}{2}(l_1+l_3+k+3)+p+{\rm v}+t}     }
L_{\frac{1}{2}(l_1+l_3+k)+p+{\rm v}+t}^{\frac{1}{2}}
\left( \frac{1+y}{1+2y}\xi^2 \right) \ ,
\label{A-anal-q}
\end{eqnarray}
}
where $\xi = qb/2$ and $y=z b^2$ ($z=\beta,\ 2\beta$).
The lower index of the Laguerre polynomial of equation (\ref{A-anal-q})
is an integer because $l_1+l_3+k$ is an even integer.

The above expression of
$\tilde{A}_{n_1l_1n_2l_2}^{n_3l_3n_4l_4k}(q,z)$ is of the form:
$f(\xi^2) \exp [- \frac{1+y}{1+2y}\xi^2]$, where
$f(\xi^2)$ is a polynomial of $\xi^2$.
The substitution of the expression of
$\tilde{A}_{n_1l_1n_2l_2}^{n_3l_3n_4l_4k}(q,z)$ to the expression of
$\tilde{O}_{22}(q,z)$, given by equation (\ref{O22-A}), leads to the
analytical expression of the two--body term of the form factor. This
expression, for the case of s--p and s--d shell nuclei, is again of the
form $\tilde{f}(\xi^2) \exp [-\frac{1+y}{1+2y}\xi^2]$ where now
$\tilde{f}(\xi^2)$ is a polynomial similar to that of
equation (\ref{f-xi2}).

From the analytical expression of the form factor,
the analytical expression of the mean square radius of a nucleus,
which is the coefficient of $-q^2/6$, can
be found. This expression is of the form:
\begin{equation}
\langle r^{2}\rangle=N [ \langle r^{2}\rangle_1
-2 \langle r^{2}(\beta b^2)\rangle_{22}
+\langle r^{2}(2 \beta b^2)\rangle_{22} ]\ ,
\label{r2}
\end{equation}
where
\begin{equation}
\langle r^{2}\rangle_1 =
[6(\eta_{1s}+\eta_{2s}) +30 \eta_{1p} +70\eta_{1d}] b^2 \ ,
\label{r2-1}
\end{equation}
and
\begin{equation}
\langle r^{2}(y)\rangle_{22} = b^2 (1+2y)^{-13/2}
\sum_{k=0}^{5} a_k y^k \ , \quad y=\beta b^2, \ 2 \beta b^2 \ .
\label{r2-1-22}
\end{equation}

The coefficients $a_k$ depend on the occupation
probabilities of the various states. If we expand, the right hand
side of equation (\ref{r2-1-22}), in powers of $1/y$
($y=\beta b^2 > 1$) and
keep powers of $\beta b^2$ up to $(\beta b^2)^{-3/2}$, an
approximate expression of the contribution of the two--body term to the
radius takes the simple form:
\begin{equation}
\langle r^{2}(y)\rangle_{22} = C b^2 (\beta b^2)^{-3/2}\ ,
\label{approx-r2}
\end{equation}
where the coefficient $C$ depends on the occupation probabilities
of the various states.

The analytical expressions of the form factor and the density, which
were found previously, will be used in section IV for the
fit of the theoretical charge form factors to the experimental ones
and for the calculations of the charge density distributions
for various N=Z (s--p and s--d shell) nuclei.

\section{RESULTS AND DISCUSSION \label{sec-4} }

The calculations of the charge form factors for various s--p and s--d
shell nuclei, with  N$=$Z, have been carried out on the basis of
equations (\ref{Fch-1}) and (\ref{Fp-3}) and the analytical expressions
of the one-- and two--body terms which were given in section III.
Two cases have been examined, named Case 1 and Case 2, which
correspond  to the analytical calculations with HO wave functions
without and with SRC respectively.
In Case 1 there is one free parameter, the HO parameter $b$, while in Case 2
there are two free parameters, the parameter $b$ and the correlation
parameter $\beta$. The parameters, in both cases, have been determined,
for each nucleus separately, by least squares fit to the experimental
$F_{ch}(q)$.

The best fit values of the parameters as well as of the values of $\chi ^2$,
\begin{equation}
\chi ^2=\frac{1}{K}\sum_{i=1}^{K}
\left [(F_i(the)-F_i(exp))/\Delta F_i(exp) \right] ^2 \ ,
\end{equation}
are displayed in Table I. In the same table the calculated root mean square
(RMS) charge radii $\langle r_{ch}^2 \rangle ^{1/2}$
and the contribution of the SRC to them,
\begin{equation}
\langle r^2 \rangle _2 =\langle r_{ch}^2 \rangle -
\langle r_{ch}^2 \rangle _1 /A \ ,
\label{r2-cor}
\end{equation}
are displayed and compared with the corresponding experimental RMS radii.
It is noted that $\langle r^2 \rangle _2$ is independent from the
centre--of--mass correction and finite proton size.

The experimental and the theoretical $F_{ch}(q)$,  for the various cases,
for the closed shell nuclei: $^{4}$He, $^{16}$O and $^{40}$Ca are
shown in Figure 1 while for the open shell nuclei: $^{12}$C, $^{24}$Mg,
$^{28}$Si and $^{32}$S, are shown in Figure 2.

From the values of $\chi ^2$, which have been found in Cases 1 and 2
(see Table I) and also from Figures 1 and 2,
it can be seen that the inclusion of the correlations improves the
fit of the form factor of all the nuclei we have examined.
Almost all the diffraction minima which are known from the experimental data
are reproduced in the correct place.
There is a disagreement in the fit of the form factor of the open shell
nuclei $^{24}$Mg, $^{28}$Si and $^{32}$S for $q\approx 3.5$ fm$^{-1}$
where it seems that there is a third diffraction minimum in the
experimental data, which cannot be reproduced in both cases.

It is seen from Table I that the parameter $b$ has the same
behavior as function of the mass number A in the HO and the correlated
model, while the following inequality holds:
$$
b(HO) > b(SRC) \ .
$$
This is due to the fact that the introduction of SRC tends to increase the
relative distance of the nucleons i.e. the size of the nucleus, while the
parameter $b$, which is (on the average) proportional to the (experimentally
fixed) radius of the nucleus, should become smaller. Such a behavior should
be expected in view also of relations (\ref{r2-1}) and (\ref{approx-r2}).

It is also noted that the difference:
\[
\Delta b =b(HO) - b(SRC) \ ,
\]
is almost constant for the open shell nuclei and it is larger for the closed
shell nuclei $^{4}$He, $^{16}$O and $^{40}$Ca.
This can also be seen  from figure 3 where the values of $\Delta b$
versus the mass number A has been plotted.
The behavior of $\Delta b$ as function of A indicates that the SRC are
stronger for the closed shell nuclei than in the open shell ones.

In Figure 4 the values of the correlation parameter $\beta$ versus the
mass number A have been plotted.
From this figure it is seen that the parameter $\beta$ is almost
constant for $^{4}$He, $^{16}$O, and $^{40}$Ca and takes larger
values (less correlated systems) in the open shell nuclei.

The behavior of the two parameters, $b$ and $\beta$, indicates that
there should be a shell effect in the case of closed shell nuclei.
That is, there is a shell effect not only on the values of the harmonic
oscillator spacing $\hbar \omega$, as has been noted in Refs.
\cite{Grypeos,Lalazi}
but also on the values of the correlation parameter $\beta$.

In the above analysis, the nuclei $^{24}$Mg, $^{28}$Si and $^{32}$S
were treated as 1d shell nuclei.
We have also considered the Case 2$^*$ in which the occupation probability
$\eta_{2s}$ of the nuclei $^{24}$Mg, $^{28}$Si and $^{32}$S is taken to
be a free parameter besides the other two parameters $b$ and $\beta$.
We found that while the $\chi^2$ values become better, comparing to those
of Case 2, the third diffraction minimum is not reproduced either and the
behavior  of the parameters $b$ and $\beta$ as functions of the mass number
$A$ remains the same. The results in this case are shown in
Table I and in figure 2. The values of the occupation probability
$\eta_{2s}$ of the above mentioned three nuclei are:
0.0355, 0.0245 and 0.2945 respectively, while the corresponding values
of $\eta_{1d}$, which can be found from the values of $\eta_{2s}$ through
the relation:
\[
\eta_{1d} = [(Z-8) - 2 \eta_{2s}]/10
\]
are: 0.3929, 0.5951 and 0.7411 respectively.
The partial occupancy of the state 2s for the nucleus $^{32}$S has as
result to increase the central part of the charge density significantly.

In Figures 5 and 6 the charge densities, $\rho_{ch}(r)$ (normalized to Z),
of the above mentioned nuclei for the various cases of Table I are shown and
compared with the charge densities of Ref. \cite{DeVries82}.
In the same figures the contribution of the SRC to $\rho_{ch}(r)$,
\begin{equation}
\rho_{2,ch}(r)= \rho_{ch}(r) - \rho_{1,ch}(r) \ ,
\label{rho-2,ch}
\end{equation}
is shown. The introduction of short range correlations has the feature of
reducing the central part of the densities of the closed shell nuclei,
while $\rho_{2,ch}(r)$ is small and characterized by oscillations in the
case of open shell nuclei.

From the determined mass dependence of the parameters $b$ and  $\beta$,
the values of these parameters for other s--p or s--d shell nuclei can
found.
In Figures 1d and 5d the $F_{ch}(q)$ and $\rho_{ch}(r)$ of the nucleus
$^{36}$Ar, treated as an 1d closed shell nucleus, are shown.
As there are no experimental data for $F_{ch}(q)$ for high $q$ values,
the value of the parameter $\beta$ is taken to be the mean value of the
corresponding values of $^{16}$O and $^{40}$Ca, that is :
$\beta_{36}=2.2937$ fm$^{-2}$, while the parameter $b$ is determined
assuming that
\[
\Delta b_{36}=\Delta b_{40}\ ,
\]
where
\[
\Delta b_{A}= b_{A}(HO)- b_A(SRC) \ .
\]
Using the values of the parameters $b_{40}(HO)=1.9453$ fm,
$b_{40}(SRC)=1.8660$ fm  from  Table I and choosing the parameter
$b_{36}(HO)=1.8800$ fm in order to reproduce the experimental
RMS charge radius of $^{36}$Ar ($<r^2>_{exp}^{1/2}=3.327\pm 15$ fm
\cite{DeVries82})  the value $b_{36}(SRC)=1.8007$ fm was found.
These values of $\beta_{36}$ and $b_{36}(SRC)$ have been used for the
calculations of the correlated $F_{ch}(q)$ and $\rho_{ch}(r)$ of
$^{26}$Ar, which are shown in figures 1d and 5d, respectively.
The calculated RMS charge radius, $<r^2>^{1/2}=3.3343$ fm,
which was found, is within the experimental error.

The effect of SRC on the form factors has also been examined
with various Skyrme--type wave functions.
In this case, there is only one free parameter, the correlation parameter
$\beta$. The potential parameters have been adjusted in Refs.
\cite{Vauth,Beiner} in order to reproduce various physical quantities such
as RMS radii, binding energies e.t.c.. Thus, some effects of the SRC, we
would like to study, have already been averaged out. Because of that we
cannot study the effect of the SRC on the parameters of the Skyrme
interactions.
This can be done if both the Skyrme parameters and the correlation
parameter are readjusted by fit to various experimental data
including the form factors and the fit is made for many
nuclei at the same time. As this is out of the scope of the present
work, we examined the form factors of $^{16}$O and $^{40}$Ca
with various Skyrme interactions, namely SK1 to SK6 \cite{Vauth,Beiner},
without SRC, named Case 3 and with SRC, named Case 4.

In Case 3 and for $^{16}$O, only SK1 gives smaller value
for $\chi ^2$ compared with that of Case 1 (HO without SRC).
The inclusion of SRC (Case 4) to the Skyrme--type
wave functions gives better $\chi ^2$ only for the SK1, but still
this value is about 10\% larger than in Case 2 (HO with SRC) and 1\%
smaller than in Case 3. For SK2 to SK6 the correlated
parameter $\beta$ goes to very large values (very small correlations) without
improving the quality of the fit.
For $^{40}$Ca and for Case 3 all the Skyrme interactions, we have examined,
give almost the same $\chi ^2$ value which is about 20\% smaller than in
Case 1 and 10\% larger than that of Case 2, while the inclusion
of the correlations improves the quality of the fit, for all the Skyrme
interactions, not more than 2\%. See Table I and figure 1
for the results we have found with SK1.

From the results with Skyrme--type wave functions we could conclude that
even if a mean field is more realistic than the HO one, the inclusion
of the SRC does not improve the fit, at least of the $F_{ch}(q)$,
significantly. Significant improvement of the fit should be
expected if the parameters of the mean field will be determined along
with the correlation parameter. It should be noted also that, a good fit
to the experimental $F_{ch}(q)$ of $^{6}$Li, $^{12}$C and $^{16}$O
has been found by Ciofi Degli Atti et al \cite{Ciofi70}, using Woods--Saxon
wave functions, only if the correlation parameter and the
depths and radii of the wells were free parameters .

\section{SUMMARY}
In the present work, general expressions for the correlated charge
form factors and densities have been found using the factor cluster
expansion of Clark et al. These expressions can be used, either for
analytical calculations, with HO orbitals or for numerical
calculations of $F_{ch}(q)$ and $\rho_{ch}(r)$,
with more realistic orbitals.

The systematic calculations, with HO orbitals, of these quantities, for
a number of N=Z, s--p and s--d shell nuclei, show that there is a shell
effect on the values of the HO parameter $b$ and on the correlation
parameter $\beta$.
Regarding the parameter $\beta$ it is almost constant for the closed
shell nuclei while it takes larger values for the open shell nuclei.
The mass dependence of these parameters indicates that the SRC are
stronger for the closed shell nuclei than for the open shell ones.
This dependence can also be used to find reasonable values of
these parameters for nuclei for which there are no experimental data
of the $F_{ch}(q)$.

Numerical calculations with various Skyrme--type wave functions, taken
from the literature \cite{Vauth,Beiner}, indicate that even if a mean field
is more realistic than the HO one, the introduction  of the SRC does not
improve the fit of the charge form factors significantly. The reason for
this should be that the introduction of the correlations makes a change in
the mean field in a way that there is a balance between the SRC and the
mean field, while for a given mean field there is not this flexibility.
A way to overcome this difficulty should be to readjust the parameters of
the mean field and the parameter of the correlations.

\section*{ACKNOWLEDGMENTS}

The authors would like to thank Professor M.E. Grypeos and Dr. C.P. Panos
for useful comments on the manuscript.



\begin{table}
\caption{The values of the parameters $b$ and $\beta$, of the $\chi^{2}$,
and of the RMS charge radii $\langle r_{ch}^{2}\rangle^{1/2}$:
contribution  of the one body density (column HO),
contribution of  SRC (column SRC) and of the total RMS charge radii
(column Total),
for various s--p and s--d shell nuclei, determined by fit to the
experimental $F_{ch}(q)$. Case 1  refers to the  HO form factor, Case 2 when
SRC are included. Case 2$^*$ is the case as case 2 but with the occupation
probability of the state $2s$ taken to be a free parameter. Cases 3 and 4
refer to SK1 potential with and without SRC, respectively.
The experimental RMS charge radii are from Ref. \cite{DeVries82}.}
\begin{center}
\begin{tabular}{ccccccccc}
\hline
& & &  & &  &  &  & \\
Case& Nucleus & $b$ [fm] & $\beta$ [fm$^{-2}$] & $\chi^{2}$ &
\multicolumn{4}{c}
{$\langle r_{ch}^{2}\rangle^{1/2}$ [fm]} \\
\multicolumn{5}{c}{}&\multicolumn{4}{c}{} \\
\cline{6-9}
&&&&&&&& \\
& & & & & HO & SRC & Total & Exper.\\
&&&&&&&& \\
\hline
&&&&&&&& \\
 1 & $^{4}He$ & 1.4320 & -- & 31 &1.7651 & -- & 1.7651 & 1.676(8)   \\
 2 & $^{4}He$ & 1.1732 & 2.3126 & 3.5 &1.5353 &0.5277 & 1.6234 &  \\
&&&&&&&& \\
 1 & $^{12}C$ & 1.6251 & -- & 177 &2.4901 & -- & 2.4901 & 2.471(6)    \\
 2 & $^{12}C$ & 1.5923 &3.7051 & 110 &2.4463 &0.2566 & 2.4597 &    \\
&&&&&&&& \\
 1& $^{16}O$ & 1.7610 & -- & 199 &2.7377 & -- & 2.7377 & 2.730(25)  \\
 2& $^{16}O$ & 1.6507 & 2.4747 & 120 &2.5853 &0.7070& 2.6802 &   \\
 3& $^{16}O$ & -- & &148 &2.6518 & -- &2.6518  &   \\
 4& $^{16}O$ & -- & 3.0201 & 146 &2.6518 & 0.1503&2.6561 &   \\
&&&&&&&& \\
 1 & $^{24}Mg$ & 1.8495 & -- & 188 &3.1170 & -- & 3.1170 & 3.075(15)   \\
 2 & $^{24}Mg$ & 1.8270 & 6.6112 & 161 &3.0823 &0.3009 & 3.0969 &    \\
 2$^*$& $^{24}Mg$ & 1.8315 & 7.1226 & 155 &3.0893 &0.2841 & 3.1023 &    \\
&&&&&&&& \\
 1 & $^{28}Si$ & 1.8941 & -- & 148 &3.2570 & -- & 3.2570 & 3.086(18)  \\
 2 & $^{28}Si$ & 1.8738 &8.2245 & 114 &3.2249 &0.2438& 3.2341 &   \\
 2$^*$&$^{28}$Si & 1.8743 &8.8104 & 112 &3.2260 &0.2315& 3.2343 &   \\
&&&&&&&& \\
 1 & $^{32}S$ & 2.0016 & -- & 320 &3.4830 & -- & 3.4830 & 3.248(11)  \\
 2 & $^{32}S$ & 1.9810 &9.1356 & 270 &3.4497 &0.2114& 3.4561 &   \\
 2$^*$ & $^{32}S$ & 1.9056 &15.579 & 194 &3.3284 &0.1445& 3.3315 &   \\
&&&&&&&& \\
 1 & $^{36}Ar$ & 1.8800 & --    & -- &3.3270& --    & 3.3270 & 3.327(15)   \\
 2 & $^{36}Ar$ & 1.8007 &2.2937 & -- &3.1970&0.9470 & 3.3343 &    \\
&&&&&&&& \\
 1 & $^{40}Ca$ & 1.9453 & --    & 229 &3.4668 & --   & 3.4668 & 3.479(3)   \\
 2 & $^{40}Ca$ & 1.8660 &2.1127 & 160 &3.3353 &1.1115& 3.5156 &    \\
 3 & $^{40}Ca$ & --     &       & 181 &3.4097 & --   & 3.4097 &    \\
 4 & $^{40}Ca$ & --     &2.1729 & 178 &3.4097 &0.2349& 3.4178 &    \\
&&&&&&&& \\
\hline
\end{tabular}
\end{center}
\end{table}

\begin{figure}
\label{ff1-fig}
\begin{center}
\begin{tabular}{cc}
{\psfig{figure=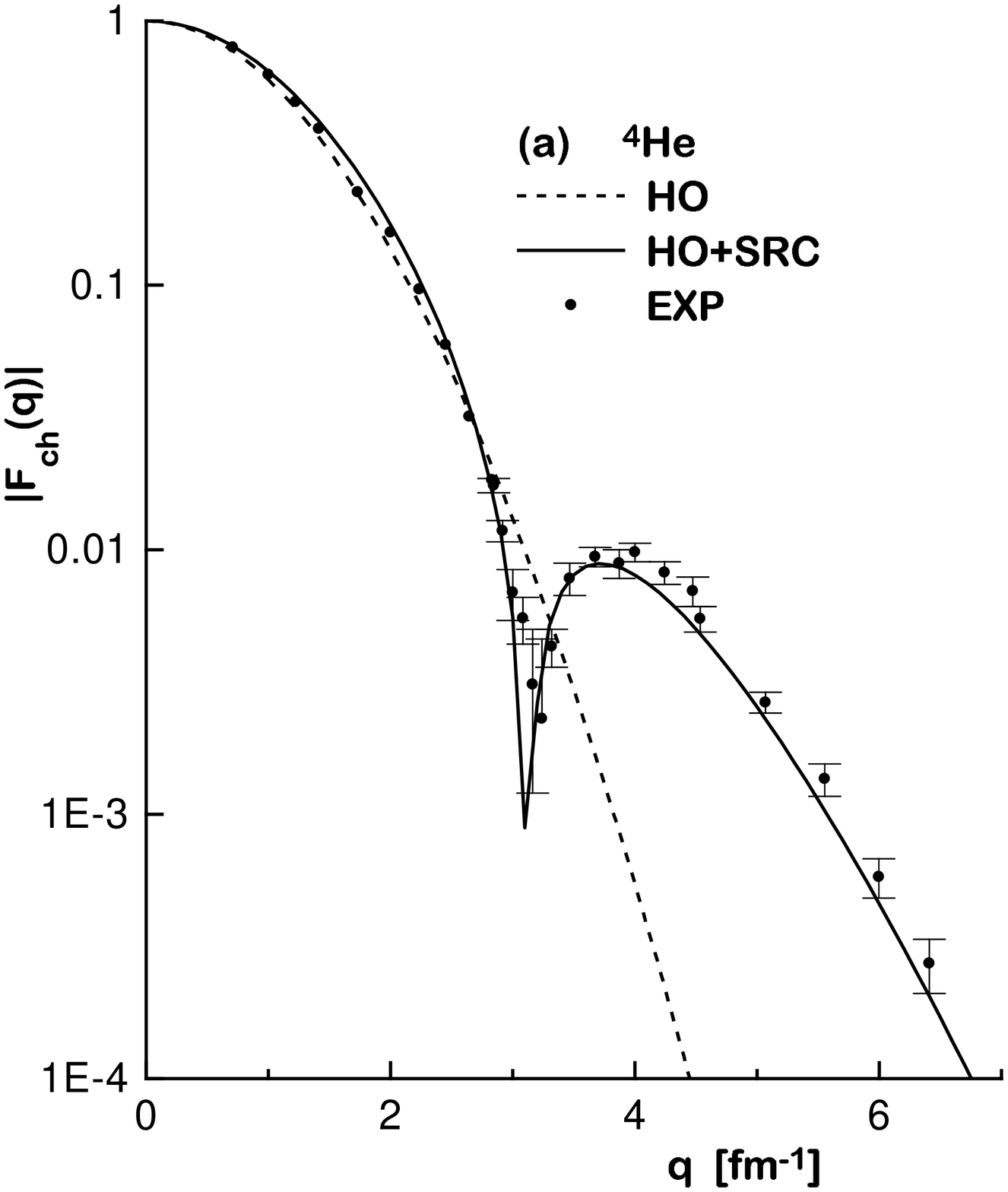,width=6.cm} } &
{\psfig{figure=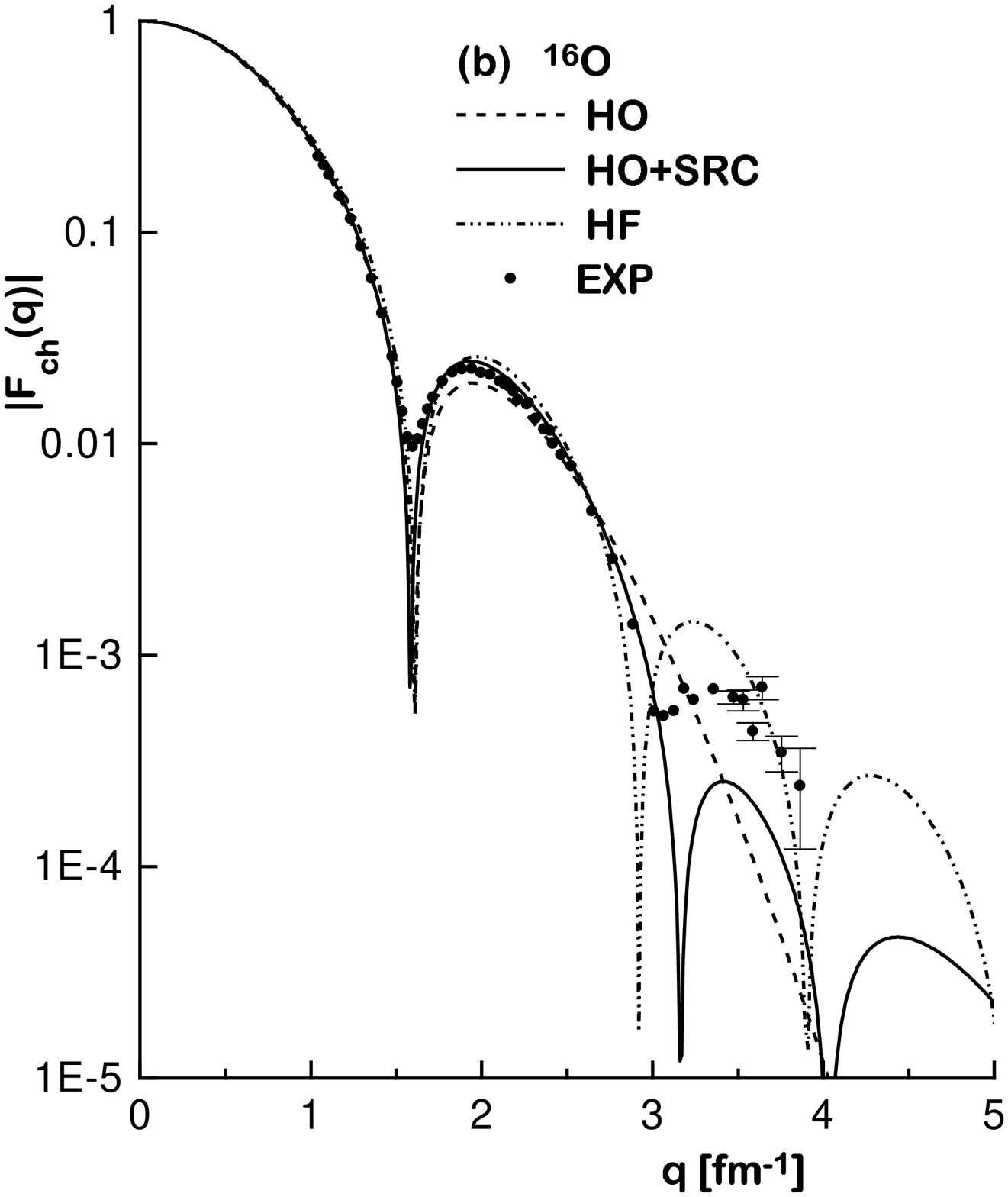,width=6.cm} }  \\
{\psfig{figure=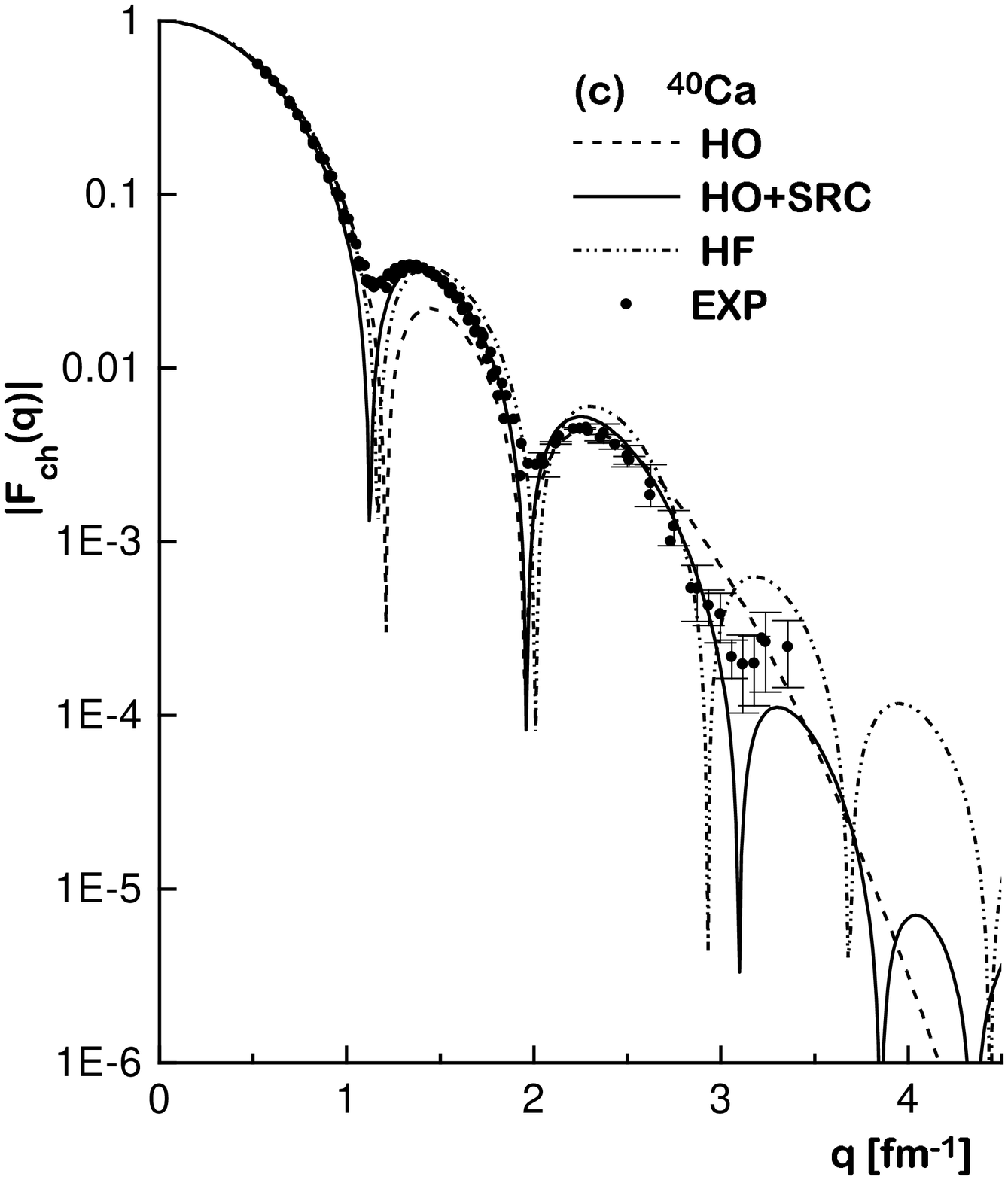,width=6.cm} } &
{\psfig{figure=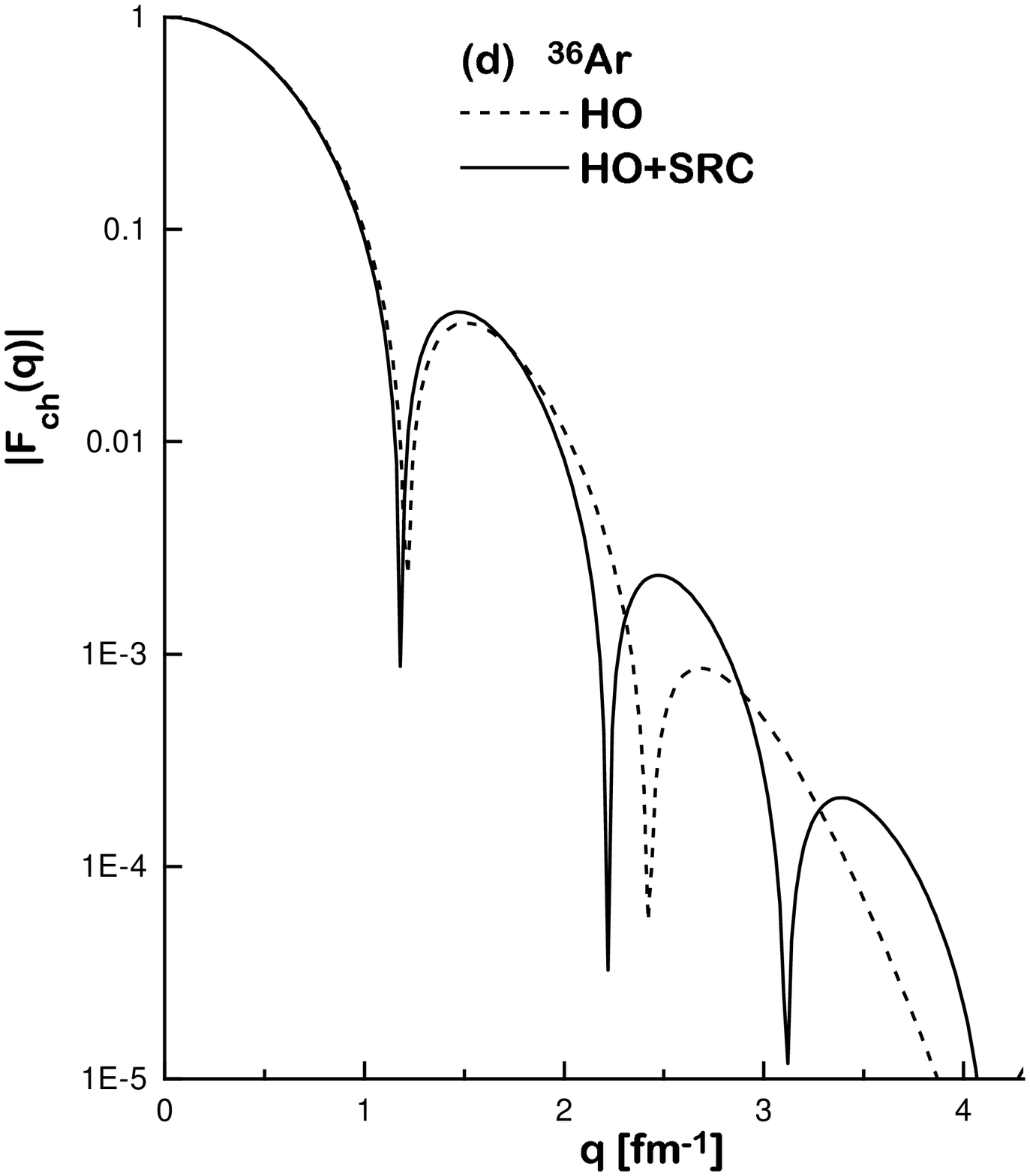,width=6.cm} }  \\
\end{tabular}
\end{center}
\caption{The charge form factor of the nuclei: $^4$He (a), $^{16}$O (b),
$^{40}$Ca (c) and $^{36}$Ar (d) for various cases. The experimental points for
$^4$He, $^{16}$O, and $^{40}$Ca are from Refs. \cite{Frosc67,Sick70,Sinha73}
respectively.}
\end{figure}

\begin{figure}
\label{ff2-fig}
\begin{center}
\begin{tabular}{cc}
{\psfig{figure=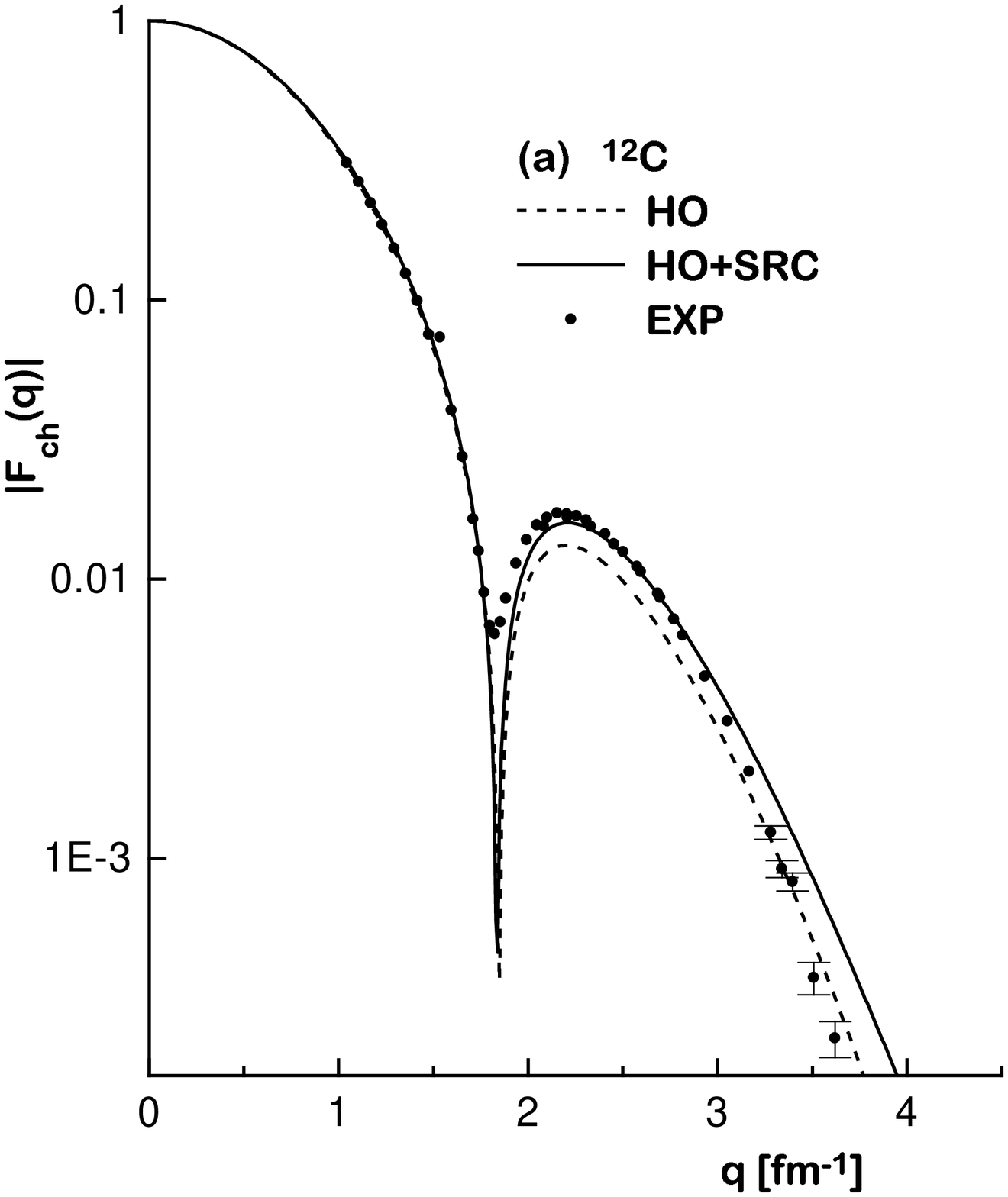,width=6.cm} } &
{\psfig{figure=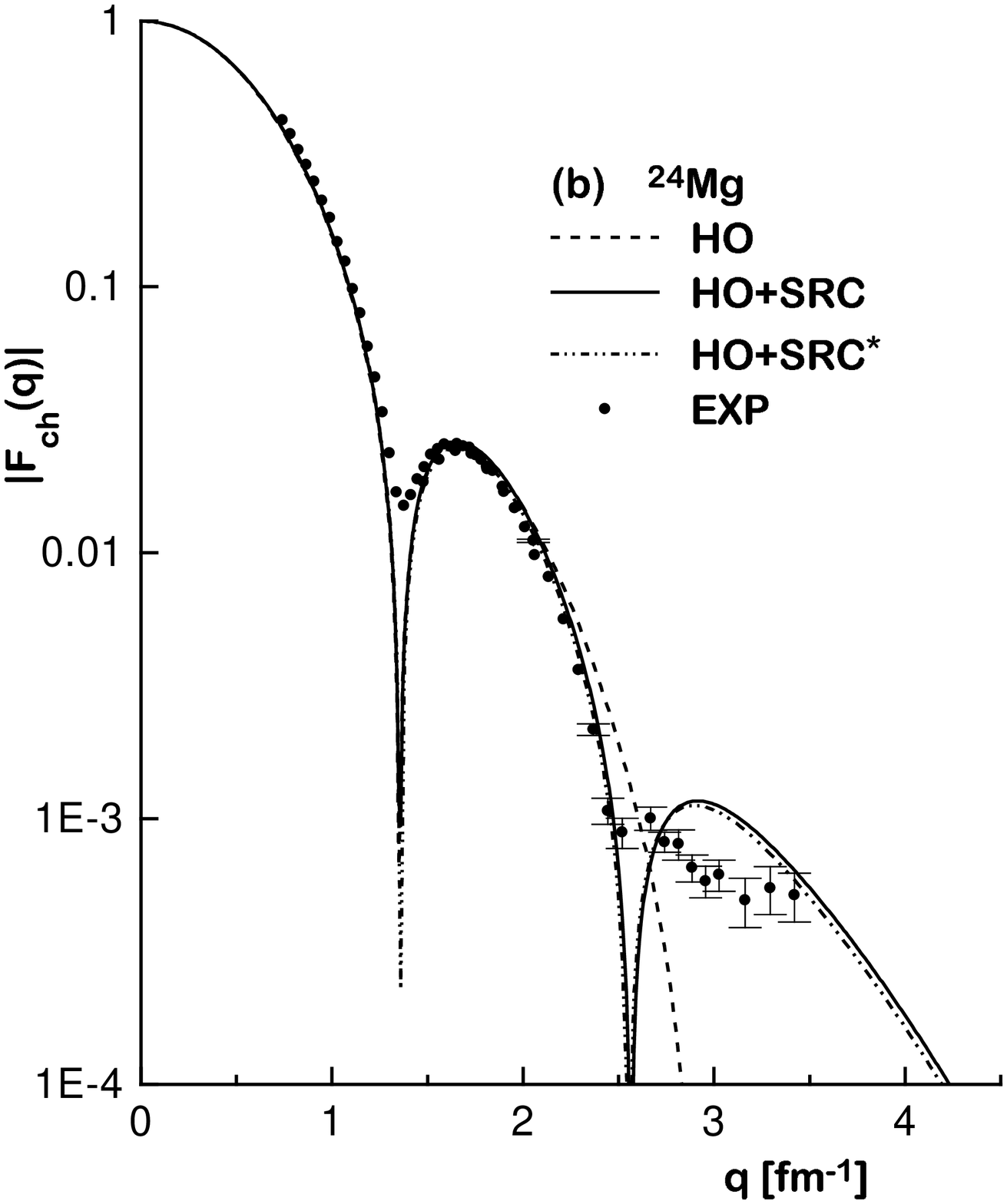,width=6.cm} }  \\
{\psfig{figure=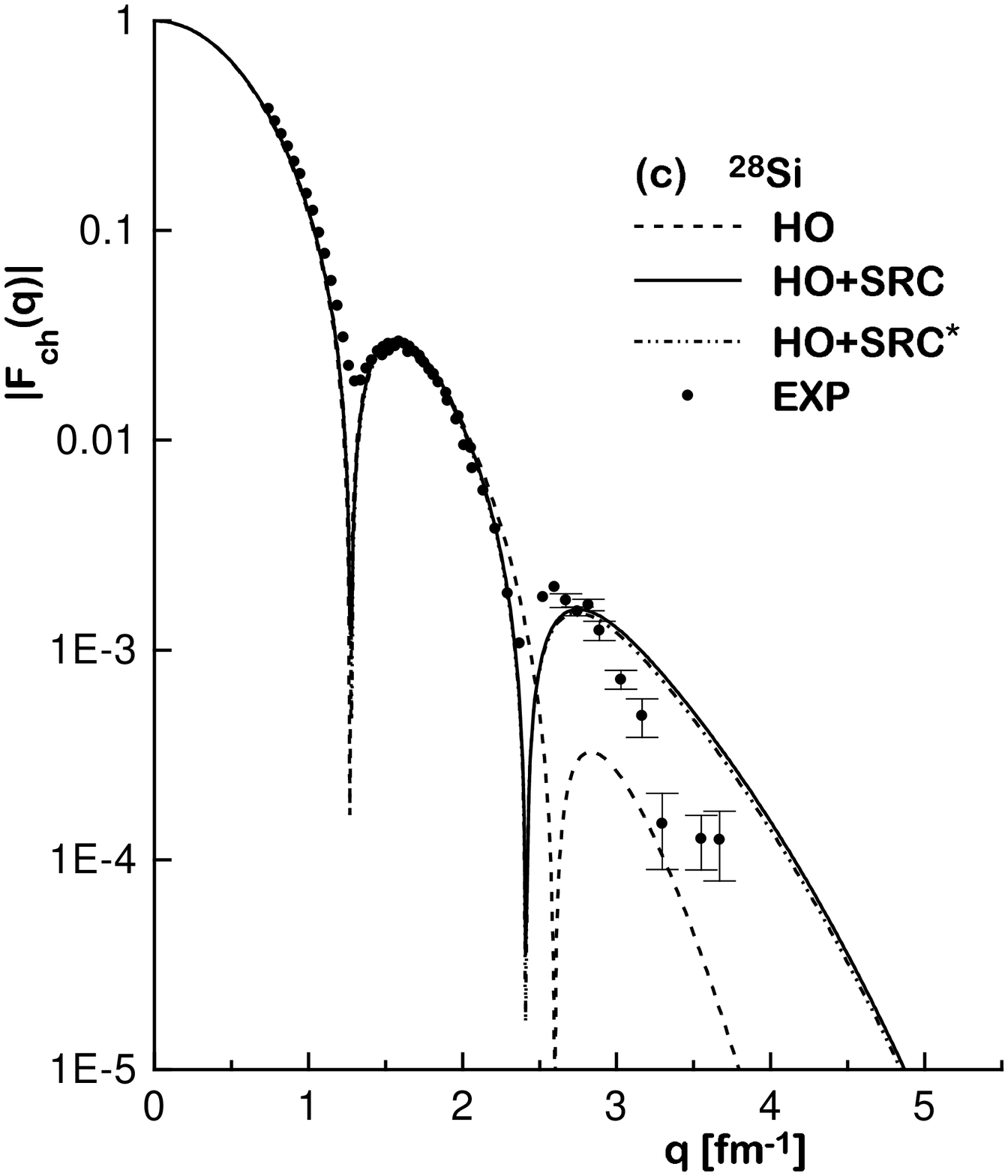,width=6.cm} } &
{\psfig{figure=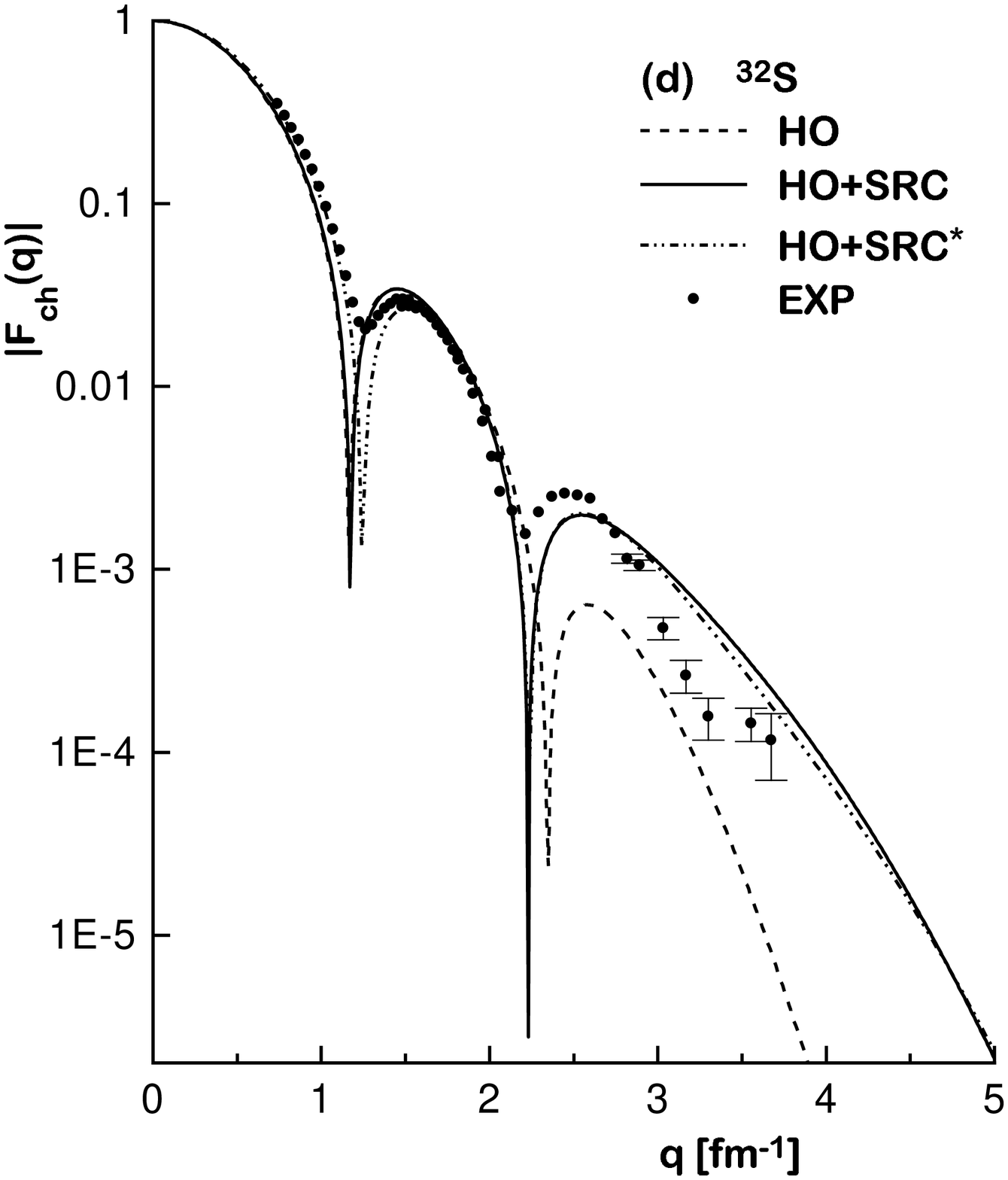,width=6.cm} }  \\
\end{tabular}
\end{center}
\caption{The charge form factor of the nuclei: $^{12}$C (a), $^{24}$Mg (b),
$^{28}$Si (c) and $^{32}$S (d) for various cases.
The case HO+SRC$^*$ corresponds to the case when the occupation
probability $\eta_{2s}$ is treating as free parameter.
The experimental points of
$^{12}$C are from Ref. \cite{Sick70} and for the other nuclei from Ref.
\cite{Li74}.}
\end{figure}

\begin{figure}
\label{b-fig}
\centerline{\psfig{figure=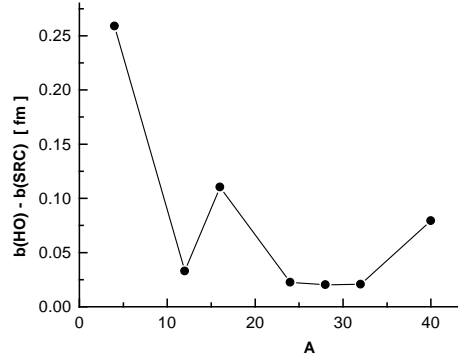,width=6.5cm} }
\caption{The difference $\Delta b= b(HO)-b(SRC)$ versus the mass number A.
$b(HO)$ and $b(SRC)$ are the HO parameter in Cases 1 (HO without SRC) and
2 (HO with SRC) respectively.}
\end{figure}

\vspace*{1cm}
\begin{figure}
\label{beta-fig}
\centerline{\psfig{figure=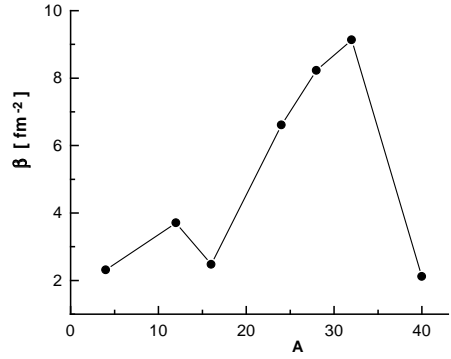,width=6.5cm} }
\caption{The correlation parameter $\beta$ versus the mass number A
in Case 2 (HO+SRC).}
\end{figure}

\begin{figure}
\label{den1-fig}
\begin{center}
\begin{tabular}{cc}
{\psfig{figure=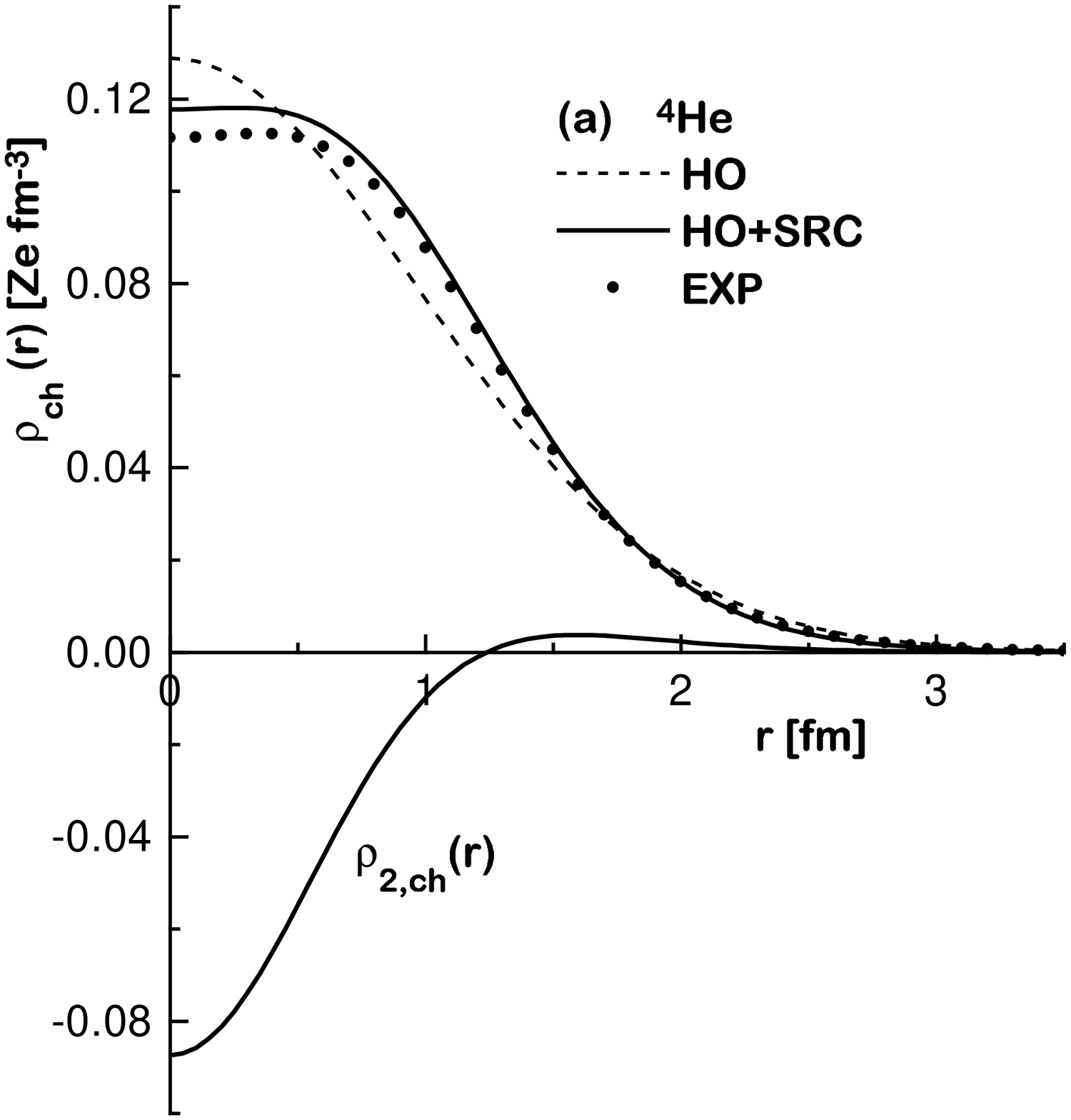,width=6.cm} } &
{\psfig{figure=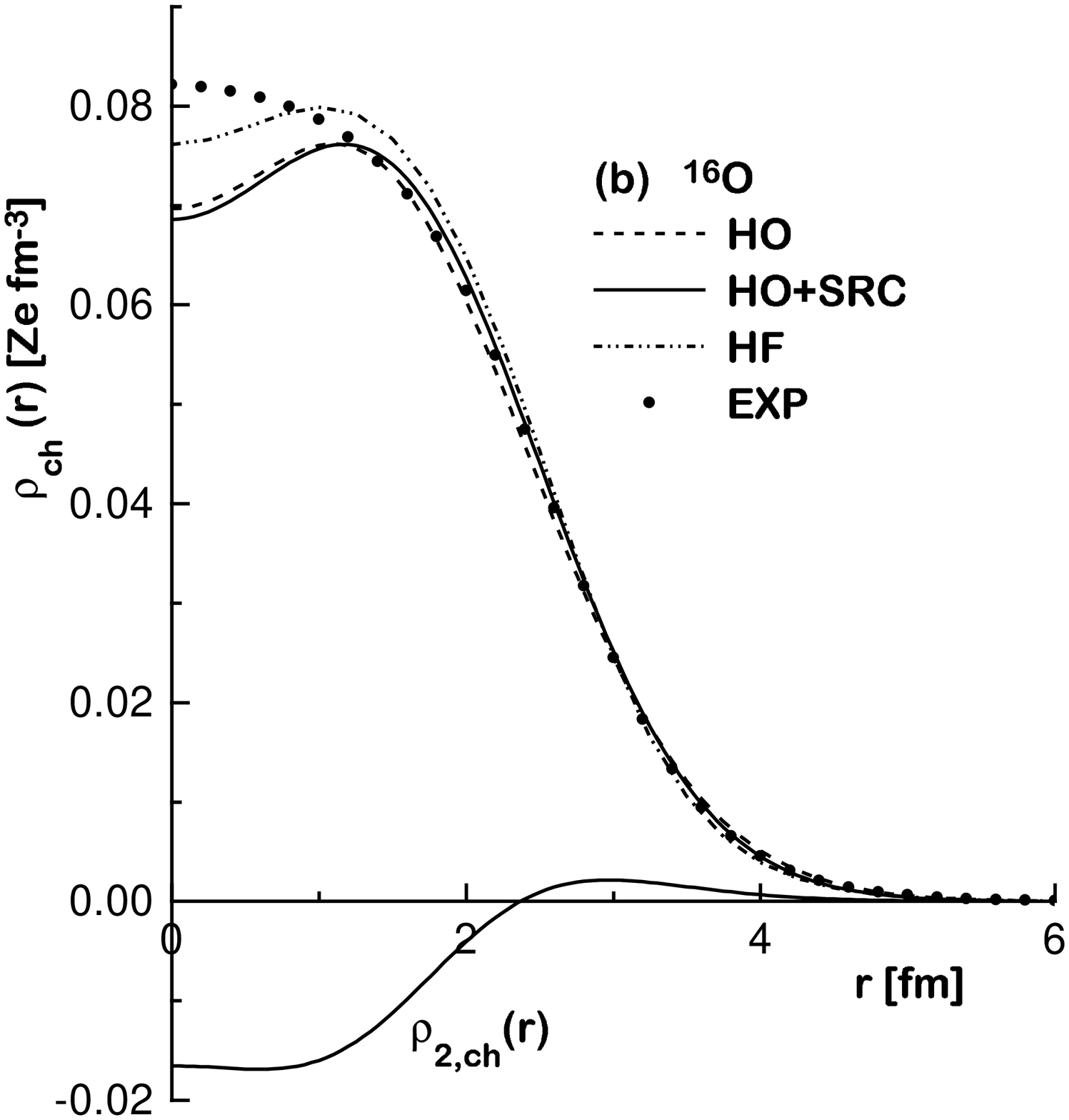,width=6.cm} }  \\
{\psfig{figure=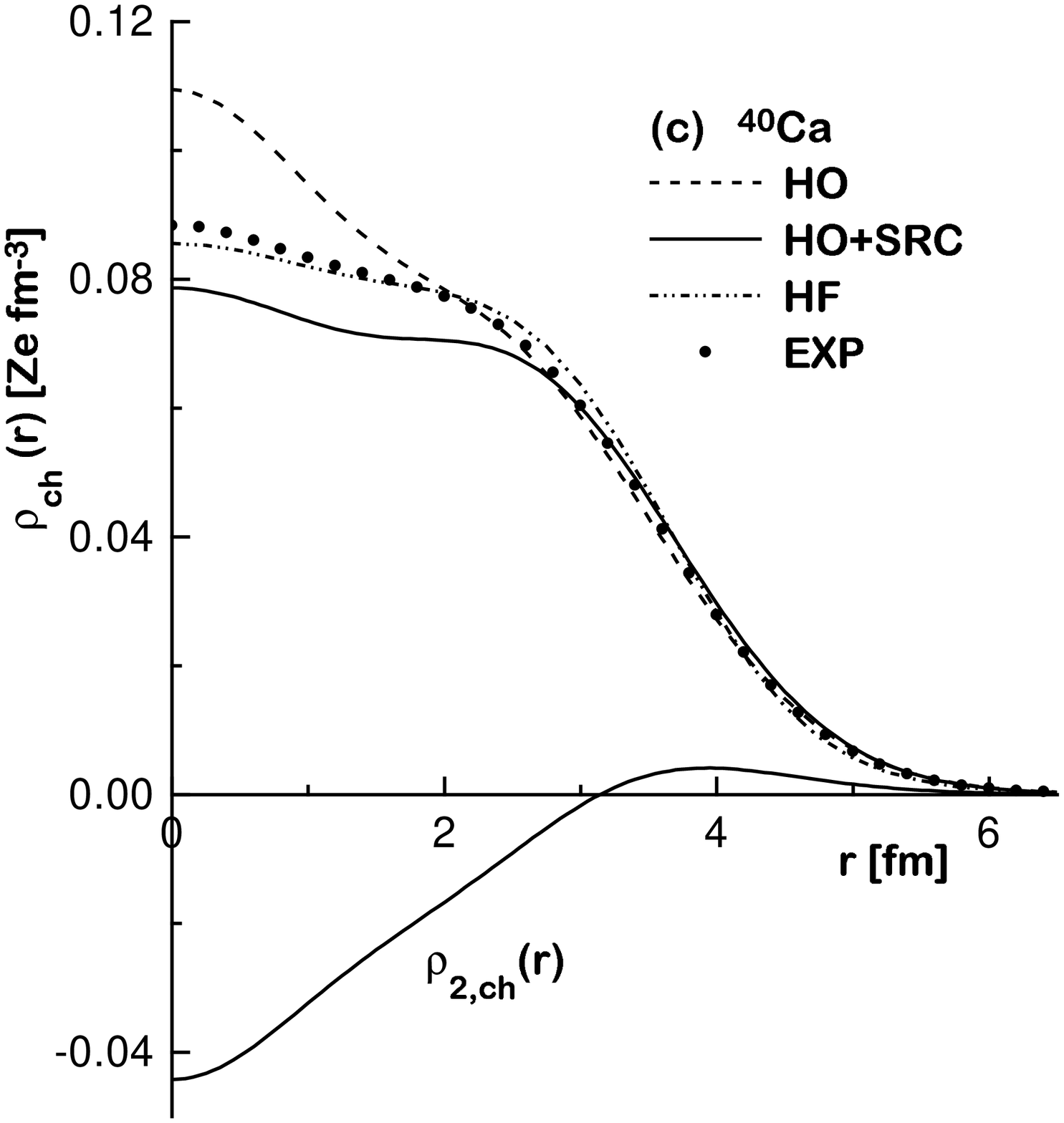,width=6.cm} } &
{\psfig{figure=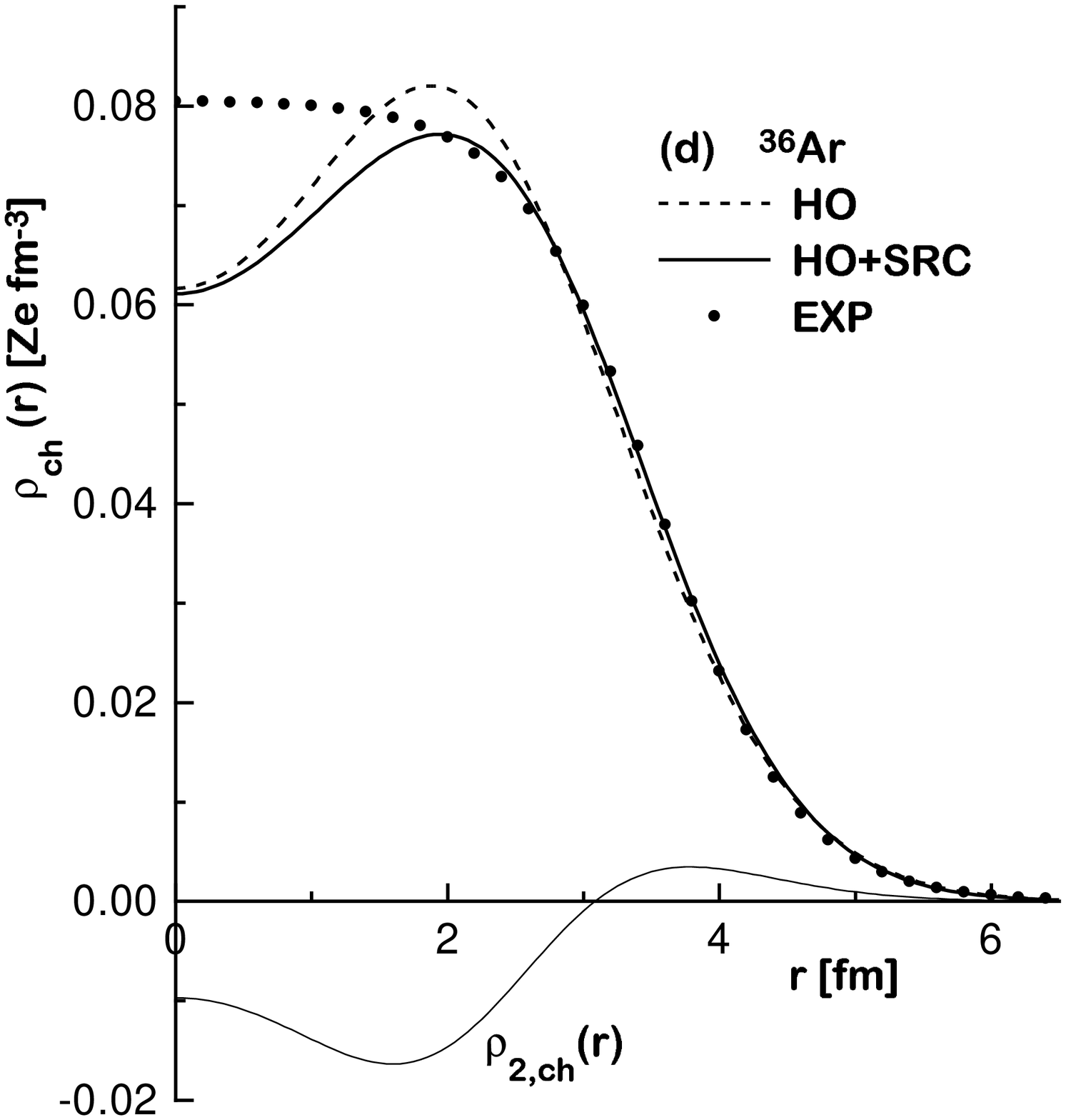,width=6.cm} }  \\
\end{tabular}
\end{center}
\caption{The charge density distribution $\rho_{ch}(r)$ and the
contribution $\rho_{2,ch}(r)$ of SRC to it of the nuclei:
$^4$He (a), $^{16}$O (b), $^{40}$Ca (c) and $^{36}$Ar (d)
for various cases.
The experimental points are from Ref. \cite{DeVries82}.}
\end{figure}

\begin{figure}
\label{den2-fig}
\begin{center}
\begin{tabular}{cc}
{\psfig{figure=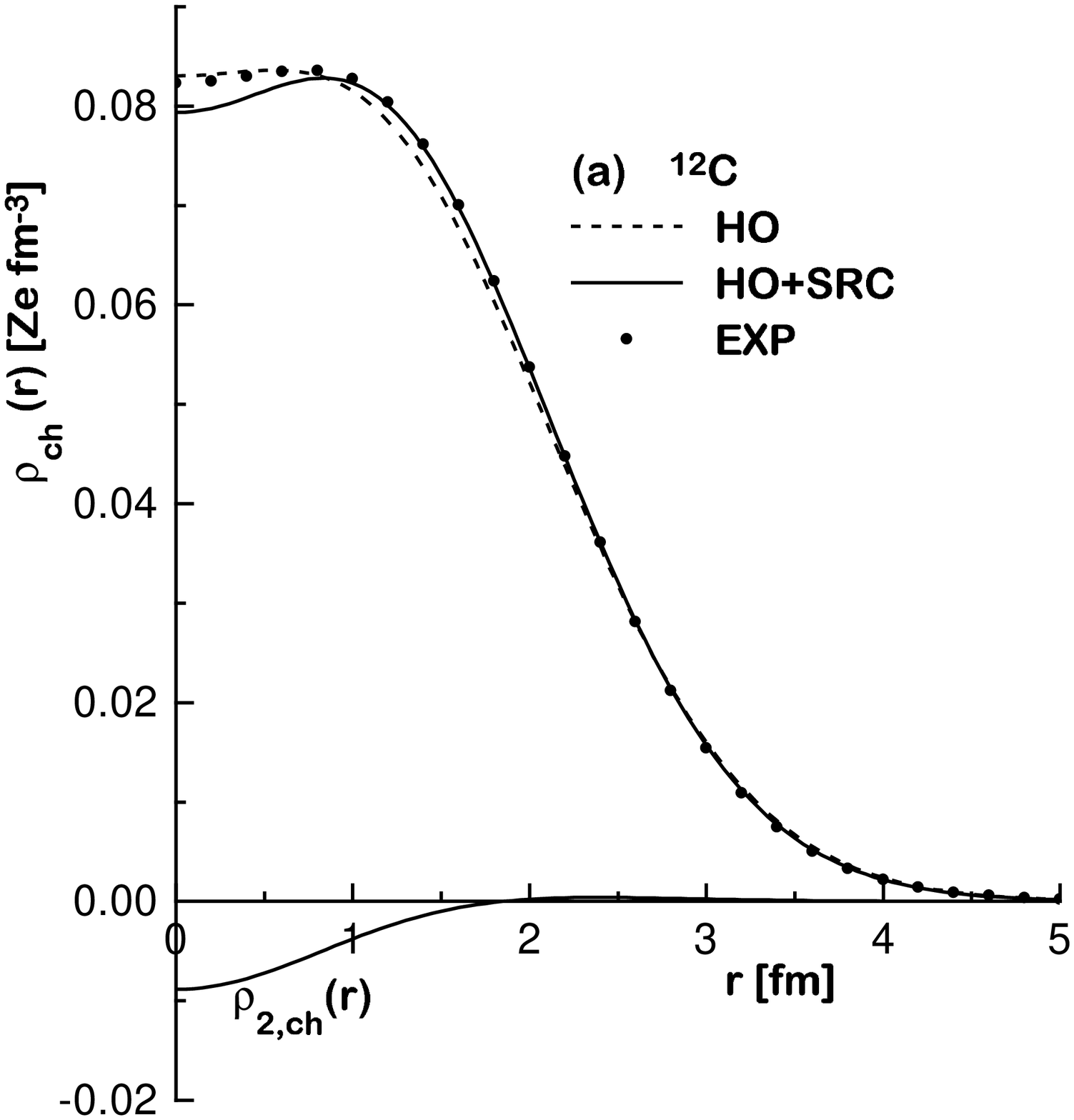,width=6.cm} } &
{\psfig{figure=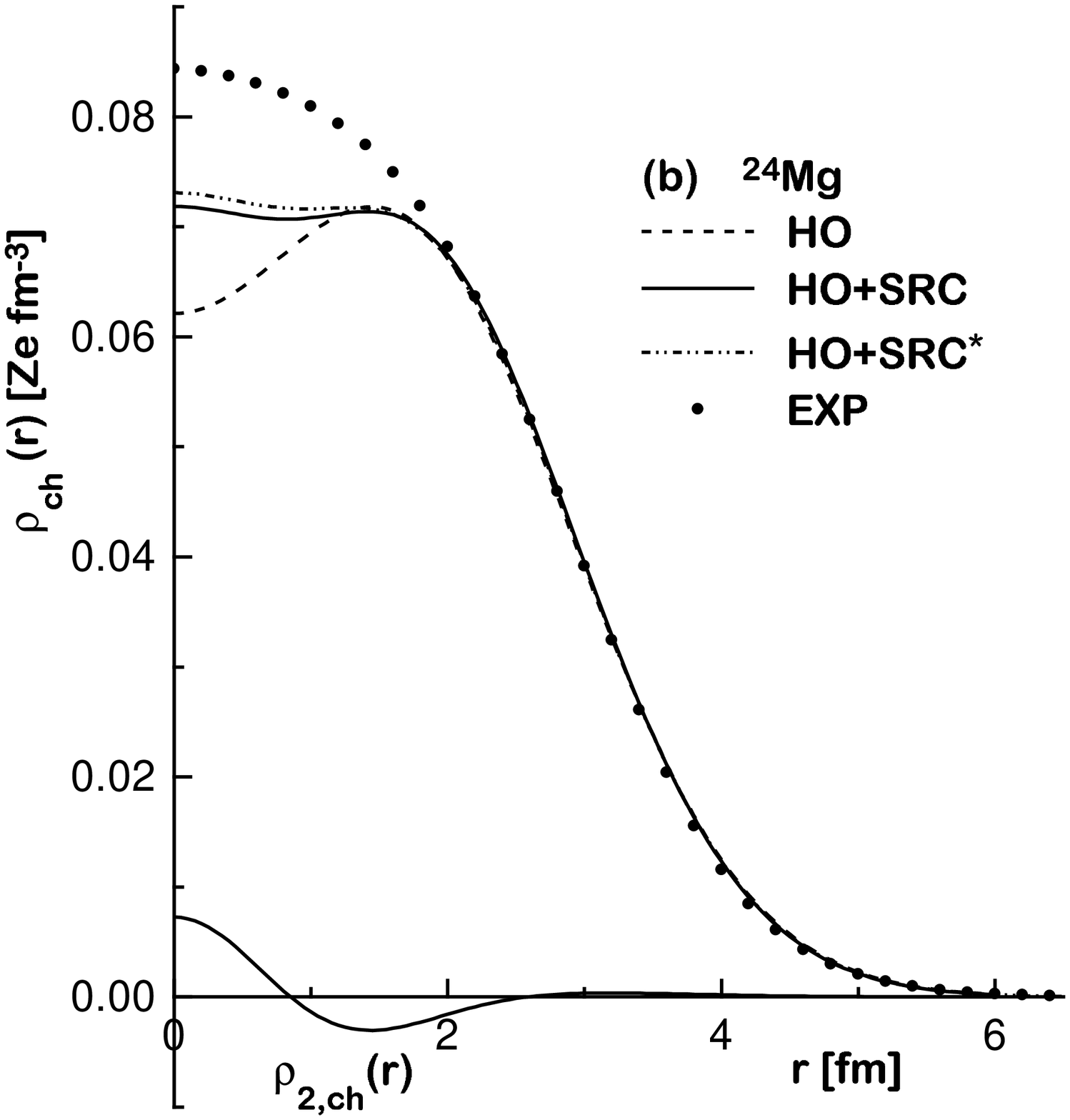,width=6.cm} }  \\
{\psfig{figure=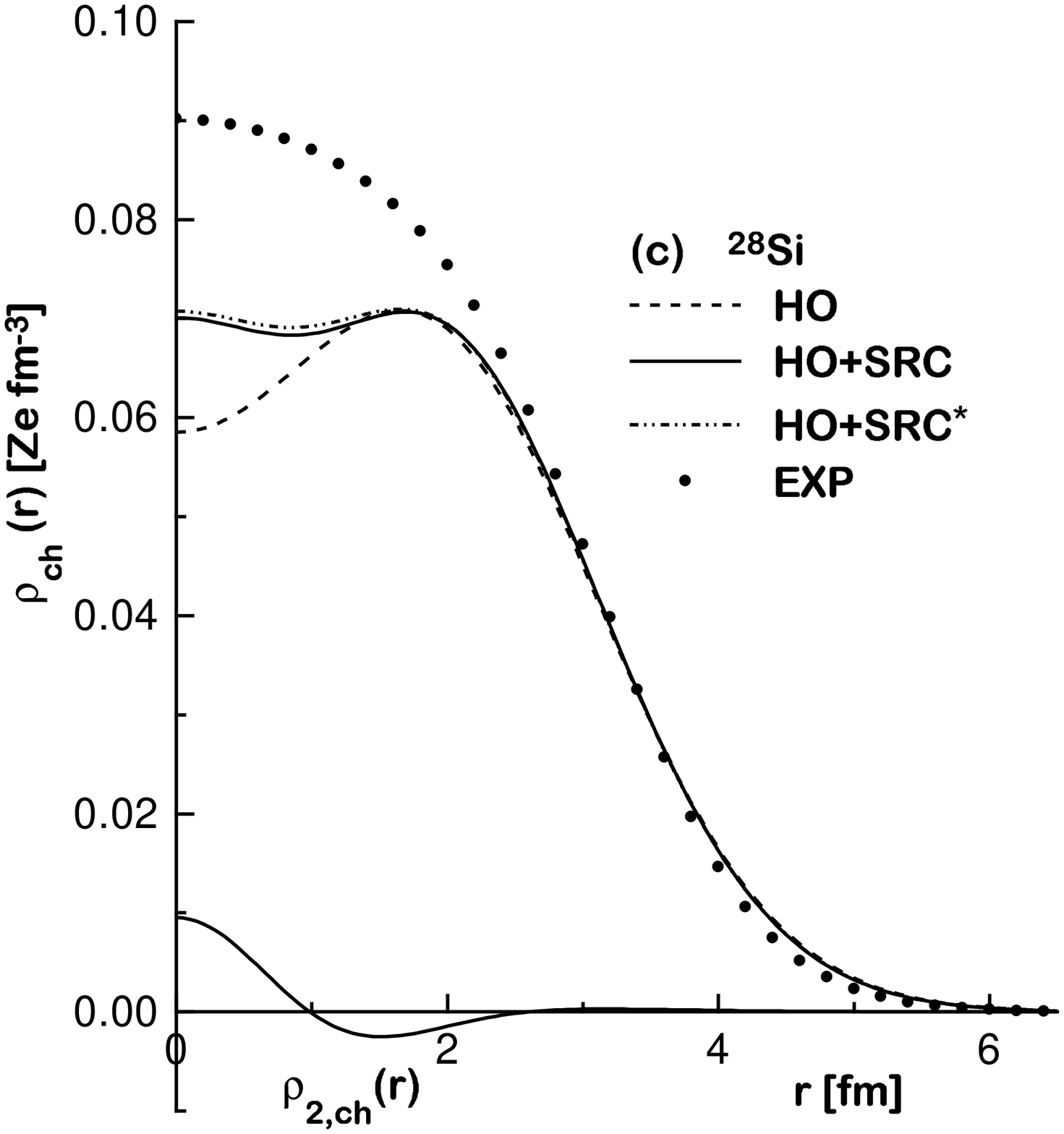,width=6.cm} } &
{\psfig{figure=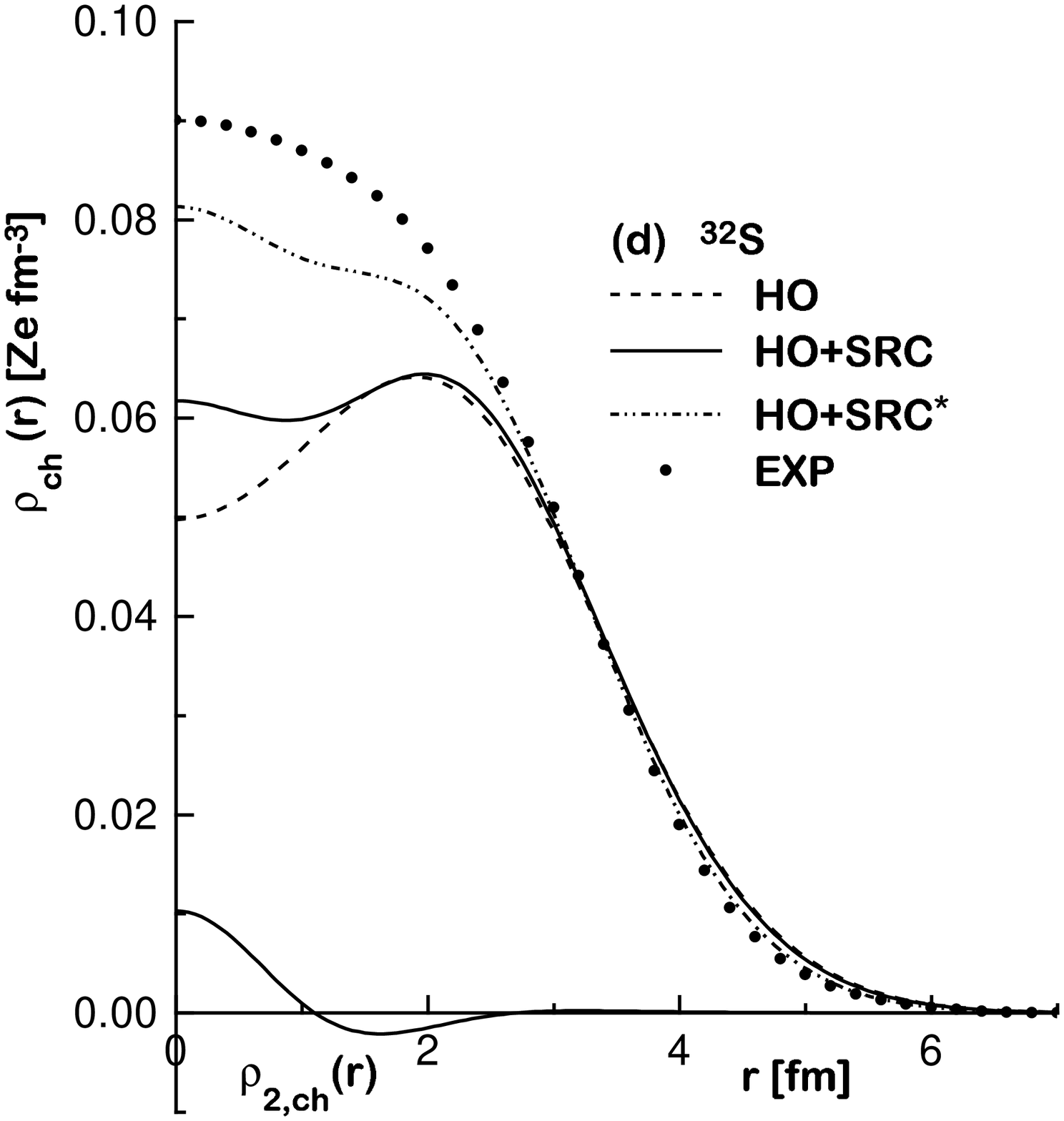,width=6.cm} }  \\
\end{tabular}
\end{center}
\caption{The charge density distribution $\rho_{ch}(r)$ and the
contribution $\rho_{2,ch}(r)$ of SRC to it of the nuclei: $^{12}$C (a),
$^{24}$Mg (b), $^{28}$Si (c) and $^{32}$S (d) for various cases.
The case HO+SRC$^*$ corresponds to the case when the occupation
probability $\eta_{2s}$ is treated as a free parameter.
The experimental points are from Ref. \cite{DeVries82}.}
\end{figure}


\begin{thebibliography}{qq}
\bibitem{Elton67}
\begin{description}
\item[\tt a.] L.R.B. Elton, {\it Nuclear Sizes} (Clarendon Press, Oxford
1961).
\item[\tt b.] H. \"{U}berall, {\it Electron Scattering from Complex Nuclei}
(Academic Press, New York and London 1971).
\item[\tt c.] R.C. Barrett, and D.F. Jackson, {\it Nuclear Sizes and
Structure} (Clarendon Press, Oxford 1977).
\end{description}

\bibitem{Czyz67} W. Czyz, and L. Lesniak, Phys. Lett. B {\bf 25}, 319 (1967).
\bibitem{Khana68} F.C. Khana, Phys. Rev. Lett. {\bf20},  871 (1968).
\bibitem{Ciofi68} C. Ciofi degli Atti, Nucl. Phys. {\bf A129}, 350 (1969).
\bibitem{Ciofi70} C. Ciofi degli Atti, and N.M. Kabachnik, Phys. Rev.
C {\bf1}, 809 (1970).
\bibitem {Ripka70} G. Ripka, and J. Gillespie, Phys. Rev.
Let. {\bf 25}, 1624 (1970).

\bibitem {Gaudin71} M. Gaudin, J. Gillespie, and G. Ripka,
Nucl. Phys. {\bf A176}, 237 (1971).

\bibitem{Bohigas80} O. Bohigas, and S. Stringari, Phys. Lett. B {\bf 95},
9 (1980).
\bibitem{DalRi82} M. Dal Ri, S. Stringari, and O. Bohigas, Nucl. Phys.
{\bf A376}, 81 (1982).
\bibitem{Nassena81} H.P. Nassena, J. Phys. G {\bf14}, 927 (1981).

\bibitem{Stoitsov93}
\begin{description}
\item[\tt a.] M.V. Stoitsov, A.N. Antonov, and S.S. Dimitrova,
        Z. Phys.  A {\bf 345}, 359 (1993).
\item[\tt b.] M.V. Stoitsov, A.N.Antonov, and S.S. Dimitrova,
       Phys. Rev. C {\bf47}, 2455 (1993).
\end{description}
\bibitem{Grypeos89}
\begin{description}
\item[\tt a.] M. Grypeos, and K. Ypsilantis, J. Phys.
G {\bf 15}, 1397 (1989).
\item[\tt b.] K. Ypsilantis, and M. Grypeos, 4th Hellenic Symposium on Nuclear
              Physics, Ioannina-Greece, (1993), 99 and J. Phys. G: {\bf 21},
1701 (1995).
\end{description}
\bibitem{Brown79} B.A. Brown, S.E. Massen, and P.E. Hodgson,
   J. Phys. G {\bf5}, 1655 (1979).
\bibitem{Malaguti82} F. Malaguti, A. Uguzzoni, E. Verodini and P.E. Hodgson,
Rivista Nuovo Cimento {\bf 5}, 1 (1982).
\bibitem{Gulkarov}
\begin{description}
\item[\tt a.] I.S. Gul'karov, Sov. J. Part. Nucl. {\bf19}, 149 (1988).
\item[\tt b.] I.S. Gul'karov, and B.P. Nigam, Phys. Rev. C {\bf 52}, 663 (1995).
\end{description}
\bibitem{Kosmas92} T.S. Kosmas, and J.D. Vergados, Nucl. Phys. {\bf A536},
72 (1992).

\bibitem{Massen88} S.E. Massen, H.P. Nassena, and C.P. Panos,
J. Phys. G {\bf14}, 753 (1988).
\bibitem{Massen89} S.E. Massen, and C.P. Panos , J. Phys. G {\bf15},
311 (1989).
\bibitem{Massen90} S.E. Massen, J. Phys. G {\bf16}, 1713 (1990).

\bibitem{Jastrow55} R. Jastrow, Phys. Rev. {\bf98}, 1497 (1955).
\bibitem{Clark70} J.W. Clark, and M. L. Ristig, Nuov. Cim. LXXA {\bf3}, 313 (1970).
\bibitem{Ristig-Clark} M.L. Ristig, W.J. Ter Low, and J.W. Clark, Phys. Rev.
C {\bf3}, 1504 (1971).
\bibitem{Clark79} J.W. Clark, Prog. Part. Nucl. Phys. {\bf 2}, 89 (1979).
\bibitem{Brink67} D.M. Brink, and M.E. Grypeos, Nucl. Phys. {\bf A97}, 81 (1967).
\bibitem{Chandra76} H. Chandra, and G. Sauer, Phys. Rev. C {\bf 13}, 245 (1976).

\bibitem{Tassie58} L.J. Tassie, and F.C. Barker, Phys. Rev. {\bf111}, 940 (1958).
\bibitem{Roy} R.R. Roy, and B.P. Nigam,  Nuclear Physics,
John Wiley \& Sons, Inc 1967.
\bibitem{deSalit} A. de-Shalit, and I. Talmi, {\it Nuclear Shell Theory},
Academic Press (1963).

\bibitem{Arias96}  F. Arias de Saavedra, G. Co', A. Fabrocini, and S. Fantoni,
Nucl. Phys. {\bf A605}, 359 (1996).

\bibitem{Arias97}  F. Arias de Saavedra, G. Co', and M.M. Renis
Phys. Rev. C {\bf  55}, 673 (1997).

\bibitem{Grads} I.S. Gradshteyn, and I.M. Ryzhik,
{\it Tables of Integrals, Series, and Products},
Academic Press (1980).

\bibitem{DeVries82} H. De Vries, C.W. De Jager, and C. De Vries,
Atom. Data and Nucl. Data Tables, {\bf 36}, 495 (1987).

\bibitem{Frosc67}
\begin{description}
\item[\tt a.] R. Frosch, Phys. Lett. B {\bf 37}, 140 (1971).
\item[\tt b.] R.G. Arnold, B.T. Chertok, S. Rock, W.P. Schutz,
Z.M Szalata, D. Day,
J.S. McCarthy, F. Martin, B.A. Mecking, I. Sick, and G. Tamas,
Phys. Rev. Lett. {\bf 40}, 1429 (1978).
\end{description}
\bibitem{Sick70} I. Sick, and J.S. McCarthy, Nucl.Phys. {\bf A150}, 631 (1970).

\bibitem{Sinha73} B.B. Sinha, G.A. Peterson, R.R. Whitney, I. Sick, and
J.S. McCarthy, Phys. Rev. C {\bf 7}, 1930 (1973).

\bibitem{Li74} G.C. Li, M.R. Yearian, and I. Sick, Phys. Rev. C {\bf 9},
1861 (1974).



\bibitem{Grypeos} C.B. Daskaloyannis, M.E. Grypeos, C.G. Koutroulos,
S.E. Massen, and  D.S. Saloupis, Phys. Lett. B {\bf 121}, 91 (1983).
\bibitem{Lalazi}
\begin{description}
\item[\tt a.] G.A. Lalazissis, and C.P. Panos, Phys. Rev. C {\bf 51},
1297 (1995).
\item[\tt b.] G.A. Lalazissis, and C.P. Panos, Intern. Journal of
Mod. Phys. {\bf E 5}, 664 (1996).
\end{description}
\bibitem{Vauth} D. Vautherin, and D.M. Brink, Phys. Rev. C {\bf 5}, 626 (1972).

\bibitem{Beiner} M. Beiner, H. Flocard, N. Van Giai, and P. Quentin,
Nucl. Phys. {\bf A238}, 29 (1975).

\end{thebibliography}
\end{document}